\def\Im{{\text{Im}}\,}
\def\Re{{\text{Re}}\,}

\def\pF{p_{\text{F}}}
\def\vF{v_{\text{F}}}

\def\sgn{{\text{sgn\,}}}
\def\be{\begin{equation}}
\def\ee{\end{equation}}
\def\bea{\begin{eqnarray}}
\def\eea{\end{eqnarray}}
\def\bse{\begin{subequations}}
\def\ese{\end{subequations}}

\documentclass[prb,twocolumn,amsmath,amssymb,eqsecnum,preprintnumbers]{revtex4}
\usepackage{graphicx}
\usepackage{dcolumn}
\usepackage{bm}
\usepackage{bbm}
\usepackage{color}
\usepackage{soul}
\usepackage{enumitem}
\begin{document}
\title{Fluctuating Quantum Kinetic Theory}
\author{T.R. Kirkpatrick$^1$ and D. Belitz$^{2,3}$}
\affiliation{$^{1} $Institute for Physical Science and Technology,
                    University of Maryland, College Park,
                    MD 20742, USA\\
                   $^{2}$ Department of Physics and Institute for Fundamental Science,
                    University of Oregon, Eugene, OR 97403, USA\\
                   $^{3}$ Materials Science Institute, University of Oregon, Eugene,
                    OR 97403, USA\\
                   }
\date{\today}

\begin{abstract}
We consider a quantum Langevin kinetic equation for a system of fermions. We first derive the Langevin
force noise correlation functions in Landau's Fermi-liquid kinetic theory from general
considerations. We then use the resulting equation to calculate the equilibrium dynamic structure
factor in the collisionless regime at low temperatures. The result is in agreement with the conventional 
many-body result. We then use the theory to derive both the fluctuating Navier-Stokes
equations for a quantum fluid and the fluctuating hydrodynamic equations for fermions in the presence
of quenched disorder. We also discuss the modifications needed for the fluctuating hydrodynamic equations 
to describe an electron fluid with long-ranged interactions, and we prove an H-theorem for the nonlinear 
Landau kinetic equation.
\end{abstract}

\maketitle

\section{Introduction}
\label{sec:I}

A Langevin equation is a stochastic equation of motion that describes a subset of degrees of freedom
of a complex system. Typically, the explicit degrees of freedom are collective (macroscopic) variables
that are assumed to change slowly with time compared to the remaining, microscopic, degrees of
freedom. The fast microscopic degrees of freedom are not described explicitly, but rather taken into 
account via a random force (``Langevin force'' or ``noise'') in the macroscopic equations of motion for 
the collective variables. These ideas for an effective or reduced theory for complex systems were first 
developed by Paul Langevin for the problem of Brownian motion, i.e., a classical particle coupled to a 
classical heat bath.\cite{Langevin_1908} Ever since they have been widely used to study problems in 
both classical and quantum statistical mechanics.

One generalization of Langevin's ideas was their application to classical many-body systems such as 
fluids, which act as their own heat bath. Examples are the equations of fluctuating hydrodynamics by 
Landau and Lifshitz,\cite{Landau_Lifshitz_VI_1959} and some of the models analyzed by Hohenberg 
and Halperin in their discussion of critical dynamics.\cite{Hohenberg_Halperin_1977} In these applications
the noise correlation is determined as follows: One assumes that the linearized hydrodynamic equations,
with Langevin forces added to the velocity and heat equations, correctly describe dynamic fluctuations
about an equilibrium state, and that the Langevin forces are delta-correlated in time. The amplitudes of
the noise correlations are then determined by the requirement that the dynamic equations yield the
correct equal-time correlation functions, which in classical statistical mechanics are related to 
thermodynamic quantities. The same reasoning has been used to develop a fluctuating classical kinetic 
theory.\cite{Dorfman_vanBeijeren_Kirkpatrick_2021} Here the reduced description is in terms of the 
fluctuating one-particle distribution function.

Other generalizations aimed at describing quantum mechanical systems. The first description of a single 
quantum particle coupled to a quantum-mechanical heat bath in terms of a Langevin equation was given 
by Ford, Kac, and Mazur;\cite{Ford_Kac_Mazur_1965} this work was later extended by 
others.\cite{Caldeira_Leggett_1983, Ford_Kac_1987, Ford_Lewis_O'Connell_1988, Olavo_Lapas_Figueiredo_2012}
Some key result were as follows. First, the most general quantum Langevin equation (QLE) can be realized 
with specific models. Second, the objects described by the QLE are Heisenberg operators. Third, the 
operator-valued Langevin force is not delta-correlated in time, i.e., the random force is not Markovian. 
The operator nature of the terms in the QLE naturally leads one to consider symmetrized and antisymmetrized  
correlation functions for both the physical quantities of interest and for the operator-valued Langevin force. 
This in turn is related to the fact that in the quantum limit there are two types of fundamental correlation functions: 
dynamic response functions and dynamic fluctuation functions.\cite{Landau_Lifshitz_IX_1991, Forster_1975}

This raises the question of how to apply Langevin's ideas to quantum many-body systems. Quantum
mechanically, the statics and the dynamics couple. As a result, the equal-time correlation functions in
general depend on the dynamics of the system, are not related to thermodynamic quantities, and are
in general not known. This means that an important ingredient of classical fluctuating hydrodynamics
is no longer applicable. On the other hand, the static response functions {\em are} in general related
to thermodynamic quantities, at least for long wavelengths, even in the quantum limit. This is true in
particular for the static response function that naturally occurs in quantum kinetic theory, and it is this
observation that we will exploit in this paper. In a previous paper (Ref.~\onlinecite{Belitz_Kirkpatrick_2021},
to be referred to as Paper I) we showed that Landau Fermi-liquid (LFL) theory is fully consistent with
hydrodynamics at all temperatures. The present paper relates to Paper I the same way 
Boltzmann-Langevin  theory\cite{Bixon_Zwanzig_1969, Dorfman_vanBeijeren_Kirkpatrick_2021} relates to the Boltzmann equation,
or flucuating hydrodynamics to Navier-Stokes theory. In contrast to Paper I, which solved the kinetic equation 
and explicitly determined the hydrodynamic frequencies and modes, the current paper will focus
on deriving the relevant hydrodynamic equations.
Specifically, we will use a fluctuating quantum kinetic theory for the one-particle distribution operator
to express the relevant dynamic response function in terms of known kinetic operators and an
anti-symmetrized Langevin force correlation function. We will then use the known static response
function to determine the amplitude of this noise correlation function. 
The fluctuation-dissipation theorem
then allows us to determine the symmetrized Langevin force correlation. From these two
results all of the relevant physical information can be obtained. 

The long-term goal of developing this formalism is to provide a method for describing the behavior
of quantum many-body systems in non-equilibrium states where conventional quantum many-body
theory is difficult to use. 

The outline of this paper is as follows. In Sec.~\ref{sec:II} we formulate the fluctuating quantum
kinetic theory and show how to derive the correlation functions for the operator-valued Langevin
force. In Sec.~\ref{sec:III} we consider three applications. In Sec.~\ref{subsec:III.A} we calculate 
the dynamic structure factor for a fermionic system in the collisionless regime. In Sec.~\ref{subsec:III.B} 
we derive the fluctuating Navier-Stokes equations for a quantum fluid and recover the results stated in 
Ref.~\onlinecite{Landau_Lifshitz_IX_1991}. In Sec.~\ref{subsec:III.C} we derive the corresponding
fluctuating hydrodynamic equations for a quantum system with quenched disorder. We conclude
with a discussion and some additional remarks in Sec.~\ref{sec:IV}. Five appendices deal
with technical details and some related issues. In Appendix A we list some relevant linearized
collision operators, Appendix B contains technical details of the derivation of the Langevin force
correlation, in Appendix C we prove an $H$-theorem for the nonlinear Boltzmann-Landau kinetic
equation, Appendix C lists some useful thermodynamic identities, and Appendix E sketches the
changes necessary for an application of the theory to electrons with a long-ranged Coulomb
interaction.

\section{Derivation of a fluctuating quantum kinetic theory}
\label{sec:II}

\subsection{A fluctuating Boltzmann-Landau equation}
\label{subsec:II.A}

Consider the $\mu$-space or single-particle phase space\cite{Ehrenfest_1907} spanned by the momentum
${\bm p}$ and the position ${\bm x}$ of a particle. If we perform a Fourier transform from the real-space
variable ${\bm x}$ to a momentum variable ${\bm k}$, the one-particle phase-space distribution operator, 
${\hat f}({\bm p},{\bm k})$, and its fluctuation, $\delta{\hat f}({\bm p},{\bm k})$, are defined by
\bse
\label{eqs:2.1}
\be
{\hat f}({\bm p},{\bm k}) = {\hat a}^{\dagger}_{{\bm p}-{\bm k}/2}\, {\hat a}_{{\bm p}+{\bm k}/2} = \delta_{{\bm k},0}\,f_{\text{eq}}({\bm p}) 
     + \delta{\hat f}({\bm p},{\bm k})
\label{eq:2.1a}
\ee
Here ${\hat a}^{\dagger}_{\bm p}$ and ${\hat a}_{\bm p}$ are creation and annihilation operators, respectively, 
for fermions with momentum ${\bm p}$, and
\be
f_{\text{eq}}({\bm p}) = \langle {\hat a}^{\dagger}_{\bm p}\, {\hat a}_{\bm p}\rangle = \frac{1}{\exp(\xi_{\bm p}/T) + 1}
\label{eq:2.1b}
\ee
\ese
is the equilibrium Fermi-Dirac distribution. $\langle\ldots\rangle$ denotes a quantum-mechanical 
expectation value plus a statistical mechanics average, and $\xi_p = \epsilon_p - \mu$ with
$\epsilon_p$ the equilibrium single-particle energy (which depends only on $p = \vert{\bm p}\vert$) and $\mu$ the chemical potential. 
Throughout this paper, by `particles' we mean quasiparticles in the sense of LFL theory.\cite{Landau_Lifshitz_IX_1991}
As in Paper I we consider spinless fermions for simplicity, and we use units such that $\hbar = k_{\text{B}} = 1$. We further define an 
operator-valued function $\hat\phi$ by
\bse
\label{eqs:2.2}
\be
\delta{\hat f}({\bm p},{\bm k}) = w({\bm p})\,{\hat\phi}({\bm p},{\bm k}) 
\label{eq:2.2a}
\ee
where
\bea
w({\bm p}) &=& -\partial f_{\text{eq}}({\bm p})/\partial\epsilon_p = \frac{1}{T}\,f_{\text{eq}}({\bm p})\left[1 - f_{\text{eq}}({\bm p})\right] 
\nonumber\\
&=& \frac{1}{4T \cosh^2(\xi_p/2T)}\ ,
\label{eq:2.2b}
\eea
\ese
For later reference we define a scalar product $\langle\ldots\vert\ldots\rangle$ in the space of functions of the momentum ${\bm p}$ with $w$ as the weight function:
\bse
\label{eqs:2.3}
\be
\langle A({\bm p})\vert B({\bm p})\rangle = \frac{1}{V}\sum_{\bm p} w({\bm p})\,A({\bm p})\,B({\bm p})\ ,
\label{eq:2.3a}
\ee
with $V$ the system volume.\cite{thermodynamic_limit_footnote}
The functions $A$ and $B$ can be number valued or operator valued. In particular,
\be
N_0 = \langle 1\vert 1\rangle
\label{eq:2.3b}
\ee
is the normalization of the weight function. Also useful is an average with respect to the weight function $w$
defined as\cite{average_footnote}
\be
\langle A({\bm p})\rangle_w = \langle A({\bm p})\vert 1\rangle/N_0\ .
\label{eq:2.3c}
\ee
\ese

Within LFL theory a small change $\delta{\hat f}$ of the distribution leads to a change $\delta\epsilon$ of the single-particle
energy given by\cite{Landau_Lifshitz_IX_1991, Landau_Lifshitz_X_1981}
\be
\delta{\hat\epsilon}({\bm p},{\bm x}) = \frac{1}{N_0V}\sum_{{\bm p}'} w({\bm p}') F({\bm p},{\bm p}')\,\delta{\hat f}({\bm p}',{\bm x})\ ,
\label{eq:2.4}
\ee
where $F({\bm p},{\bm p'})$ is Landau's interaction function that parameterizes the interaction between the fermions
and we consider a real-space variable ${\bm x}$ instead of the momentum variable ${\bm k}$. 
The distribution in Eq.~(\ref{eq:2.1a}) thus corresponds to a single-particle energy
\be
{\hat\epsilon}({\bm p},{\bm x}) = \epsilon_p + \delta{\hat\epsilon}({\bm p},{\bm x})\ ,
\label{eq:2.5}
\ee
which is operator valued and spatially varying. 

Going back to Fourier space, 
we assume that the time evolution of the dynamic fluctuation operator $\hat\phi$ is governed by a fluctuating
Boltzmann-Landau kinetic equation\cite{dynamics_footnote}
\bse
\label{eqs:2.6}
\be
\left[\partial_t + L_{\bm k}({\bm p})\right] {\hat\phi}({\bm p},{\bm k},t) = {\hat F}_{\text{L}}({\bm p},{\bm k},t)
\label{eq:2.6a}
\ee
or, after a temporal Fourier transform with $\omega$ as the frequency,
\be
\left[-i\omega + L_{\bm k}({\bm p})\right] {\hat\phi}({\bm p},{\bm k},\omega) = {\hat F}_{\text{L}}({\bm p},{\bm k},\omega)
\label{eq:2.6b}
\ee
Here
\be
L_{\bm k}({\bm p}) = L_{\bm k}^{(1)}({\bm p})  - \Lambda({\bm p})
\label{eq:2.6c}
\ee
is a linearized kinetic operator. It is comprised of a linearized collision operator $\Lambda$ and
\be
L_{\bm k}^{(1)}({\bm p}) = i{\bm k}\cdot{\bm v}_p\,\left[1 + I({\bm p})\right] 
\label{eq:2.6d}
\ee
where
\be
I({\bm p}) = \frac{1}{N_0 V}\sum_{{\bm p}'} w({\bm p}')\,F({\bm p}',{\bm p})\,R_{{\bm p}\to{\bm p}'}
\label{eq:2.6e}
\ee
with $R$ a replacement operator defined by $R_{{\bm p}\to{\bm p}'} A({\bm p}) = A({\bm p}')$.
${\bm v}_p = {\bm\nabla}_{\bm p}\epsilon_p$ is the quasiparticle velocity which obeys the integral equation\cite{Landau_Lifshitz_IX_1991}
\be
{\bm v}_p = \frac{1}{m}\,{\bm p} - \frac{1}{N_0 V} \sum_{{\bm p}'} w({\bm p}')\,F({\bm p},{\bm p})\,{\bm v}_{p'}\ .
\label{eq:2.6f}
\ee
\ese
where $m$ is the bare particle mass. Near the Fermi surface, $\epsilon_p = \mu + \vF^*(p - \pF)$, and hence ${\bm v}_p = {\bm p}/m^*$
where $\pF$ is the Fermi momentum and $\vF^* = \pF/m^*$ is the Fermi velocity with $m^*$ the quasiparticle effective mass.
The form of the linearized collision operator $\Lambda$ depends on the scattering processes considered. In Appendix~\ref{app:A} we
give examples of collision operators that describe fermion-fermion scattering, fermion-phonon scattering, and
fermion-impurity scattering, respectively. 

${\hat F}_{\text{L}}$ in Eqs.~(\ref{eq:2.6a}, \ref{eq:2.6b}) is the fluctuating $\mu$-space Langevin force operator. We assume that
it is Gaussian distributed with zero mean. The latter property follows because the average of ${\hat\phi}$ must be governed by the
averaged kinetic equation. The second moment, which then completely determines the Gaussian distribution, we will determined below.

\subsection{Response and correlation functions}
\label{subsec:II.B}

As mentioned in the Introduction, there are two different correlation functions for the variable of interest, which is $\hat\phi$.\cite{Forster_1975, Zubarev_1960} 
One is the $\mu$-space dynamic response function, or commutator correlation function, $\chi''$, defined by
\begin{widetext}
\bse
\label{eqs:2.7}
\be
\frac{1}{2} \Big\langle \left[{\hat\phi}({\bm p}_1,{\bm k}_1,\omega_1),{\hat\phi}({\bm p}_2,{\bm k}_2,\omega_2)\right]_{-}\Big\rangle  = 
   2\pi \delta(\omega_1 + \omega_2)\,V \delta_{{\bm k}_1 + {\bm k}_2,0}\,\chi''({\bm p}_1,{\bm p}_2;{\bm k}_1,\omega_1)\ .
\label{eq:2.7a}
\ee
The other is the $\mu$-space fluctuation function, or anticommutator correlation function, $\Phi$, defined by
\be
\frac{1}{2} \Big\langle \left[{\hat\phi}({\bm p}_1,{\bm k}_1,\omega_1),{\hat\phi}({\bm p}_2,{\bm k}_2,\omega_2)\right]_{+}\Big\rangle  = 
   2\pi \delta(\omega_1 + \omega_2)\,V \delta_{{\bm k}_1 + {\bm k}_2,0}\,\Phi({\bm p}_1,{\bm p}_2;{\bm k}_1,\omega_1)\ .
\label{eq:2.7b}
\ee
\ese
Here 
$[{\hat A},{\hat B}]_{\mp} = {\hat A} \mp {\hat B}$. 
$\chi''$ and $\Phi$ are
related by the fluctuation-dissipation theorem,\cite{Callen_Welton_1951, Forster_1975, Pines_Nozieres_1989} which in the present context takes the 
form
\be
\Phi({\bm p}_1,{\bm p}_2;{\bm k},\omega) = \coth(\omega/2T)\,\chi''({\bm p}_1,{\bm p}_2;{\bm k},\omega)\ .
\label{eq:2.8}
\ee
$\chi''$ is the spectrum of the related causal function\cite{Forster_1975}
\bse
\label{eqs:2.9}
\be
\chi({\bm p}_1,{\bm p}_2;{\bm k},z) =  \int \frac{d\omega}{\pi}\,\frac{\chi''({\bm p}_1,{\bm p}_2;{\bm k},\omega)}{\omega - z}
\label{eq:2.9a}
\ee
and can be obtained from the latter via the relation
\be
\chi''({\bm p}_1,{\bm p}_2;{\bm k},\omega) = \lim_{\epsilon\to 0} \Im \chi({\bm p}_1,{\bm p}_2;{\bm k},\omega+i\epsilon)
\label{eq:2.9b}
\ee
\ese
The static $\mu$-space response function can be written as the minus first frequency moment of $\chi''$,
\bea
\chi({\bm p}_1,{\bm p}_2;{\bm k}) &=& \chi({\bm p}_1,{\bm p}_2;{\bm k},z=0)
\nonumber\\
&=& \int \frac{d\omega}{\pi}\,\frac{\chi''({\bm p}_1,{\bm p}_2;{\bm k},\omega)}{\omega}\ .
\label{eq:2.10}
\eea

\smallskip
\subsection{The Langevin force correlation}
\label{subsec:II.C}

We are interested in the correlations of the Langevin noise operator ${\hat F}$. Temporal and spatial translational invariance
imply that they must have the form
\be
\frac{1}{2} \Big\langle \left[{\hat F}_{\text{L}}({\bm p}_1,{\bm k}_1,\omega_1), {\hat F}_{\text{L}}({\bm p}_2,{\bm k}_2,\omega_2)\right]_{\pm}\Big\rangle  
               = 2\pi \delta(\omega_1+\omega_2)\,V \delta_{{\bm k}_1+{\bm k}_2,0}\,\Psi_{\pm}({\bm p}_1,{\bm p}_2; {\bm k}_1,\omega_1)\ .
\label{eq:2.11}
\ee                    
By using the kinetic equation (\ref{eq:2.6b}) in Eq.~(\ref{eq:2.7a}) we obtain a formal relation between $\Psi_{-}$ and the dynamic response 
function,
\be
\chi''({\bm p}_1,{\bm p}_2;{\bm k},\omega)
  = \frac{1}{\left[-i\omega + L_{\bm k}({\bm p}_1)\right] \left[i\omega + L_{-{\bm k}}({\bm p}_2)\right]}\,\Psi_{-}({\bm p}_1,{\bm p}_2; {\bm k},\omega)\ .  
\label{eq:2.12}
\ee
\end{widetext}          
We now make an {\it ansatz} that assumes that $\Psi_{-}$ is linear in $\omega$,
\be
\Psi_{-}({\bm p}_1,{\bm p}_2; {\bm k},\omega) = \omega\,\psi({\bm p}_1,{\bm p}_2; {\bm k})\ .
\label{eq:2.13}
\ee
By using Eq.~(\ref{eq:2.10}) we then obtain an expression for $\Psi_{-}$ in terms of the static $\mu$-space response function,
\be
\Psi_{-}({\bm p}_1,{\bm p}_2; {\bm k},\omega) = \frac{\omega}{2T} \left[L_{\bm k}({\bm p}_1) + L_{-{\bm k}}({\bm p}_2)\right] T \chi({\bm p}_1,{\bm p}_2;{\bm k})\ .
\label{eq:2.14}
\ee
This result depends on the collision operator via the kinetic operator $L_{\bm k}({\bm p})$, Eq.~(\ref{eq:2.6c}).
The static response function does not depend on the collision operator, see Appendix~\ref{app:B}. 
As a result, the streaming and interaction terms in $L_{\bm k}({\bm p})$ do not contribute, and our final result is
\bse
\label{eqs:2.15}
\be
\Psi_{-}({\bm p}_1,{\bm p}_2; {\bm k},\omega) = \frac{-\omega}{2T} \left[\Lambda({\bm p}_1) + \Lambda({\bm p}_2)\right] T \chi({\bm p}_1,{\bm p}_2;{\bm k})\ ,
\label{eq:2.15a}
\ee
where $\chi$ is the solution of the integral equation
\bea
\chi({\bm p}_1,{\bm p}_2;{\bm k}) + \frac{1}{N_0V} \sum_{{\bm p}'} w({\bm p}')\,F({\bm p}_1,{\bm p}')\,\chi({\bm p}',{\bm p}_2;{\bm k}) 
     &=&
\nonumber\\
&& \hskip -110pt = V \delta_{{\bm p}_1,{\bm p}_2}\,\frac{1}{w({\bm p}_1)}\ .
\label{eq:2.15b}
\eea
See Appendix~\ref{app:B} for a derivation of these results.
Equation~(\ref{eq:2.15b}) determines the static response function. 
Finally, Eq.~(\ref{eq:2.8}) provides a corresponding expression for $\Psi_{+}$,
\be
\Psi_{+}({\bm p}_1,{\bm p}_2; {\bm k},\omega) = \coth\left(\frac{\omega}{2T}\right) \Psi_{-}({\bm p}_1,{\bm p}_2; {\bm k},\omega)
\label{eq:2.15c}
\ee
\ese
The latter result gives an {\it a posteriori} motivation for the ansatz (\ref{eq:2.13}): In the high-temperature limit, where $(\omega/2T)\coth(\omega/2T) \to 1$,
Eq.~(\ref{eq:2.15c}) is structurally identical to the corresponding result for the classical fluctuating Boltzmann equation, see Chapter 8 in 
Ref.~\onlinecite{Dorfman_vanBeijeren_Kirkpatrick_2021}.

For explicit calculations within LFL theory an explicit interaction model is often used that performs an angular-momentum expansion of the interaction function
$F({\bm p}_1,{\bm p}_2)$, truncated at first order,
\bse
\label{eqs:2.16}
\be
F({\bm p}_1,{\bm p}_2) = F_0 + F_1\, \frac{{\bm p}_1\cdot{\bm p}_2}{\langle {\bm p}^2\rangle_w}\ .
\label{eq:2.16a}
\ee
The operator $I({\bm p})$ from Eq.~(\ref{eq:2.6e}) then can be written
\be
I({\bm p}) = \frac{F_0}{\langle 1\vert 1\rangle}\,\vert 1\rangle\langle 1\vert +  \frac{F_1}{\langle {\bm p}\vert{\bm p}\rangle}\,\vert {\bm p}\rangle\langle {\bm p}\vert\ ,
\label{eq:2.16b}
\ee
\ese
the integral equation (\ref{eq:2.15b}) becomes separable,\cite{Landau_parameters_footnote} and we have
\bea
\chi({\bm p}_1,{\bm p}_2;{\bm k}) &=& V \delta_{{\bm p}_1,{\bm p}_2}\,\frac{1}{w({\bm p}_1)} - \frac{1}{N_0V}\,\frac{F_0}{1+F_0}
\nonumber\\
&&\hskip 10pt  - \frac{1}{N_0V}\, \frac{{\bm p}_1\cdot{\bm p}_2}{\langle{\bm p}^2\rangle_w}\,
\frac{F_1}{1+F_1/3}\ .
\label{eq:2.17}
\eea
This yields the correct result for the static density response function, viz., $N_0/(1+F_0) = (\partial n/\partial\mu)_{T,V}$, and the
static momentum response function, $nm^*(1+F_1/3) = nm$. Note that in general Eq.~(\ref{eq:2.17}) implies nontrivial
correlations in momentum space. These do not occur in classical fluids in equlibrium, where only the $ \delta_{{\bm p}_1,{\bm p}_2}$
term is present. If the collision operator conserves both particle number and momentum (as is the case, e.g., if the only collisions considered
are fermion-fermion collisions), then neither of the Landau parameters $F_0$ and $F_1$ contribute and we have
\be
\Psi_{-}({\bm p}_1,{\bm p}_2; {\bm k},\omega) = \frac{-\omega}{2T} \left[\Lambda({\bm p}_1) + \Lambda({\bm p}_2)\right]\,
      V \delta_{{\bm p}_1,{\bm p}_2}\,\frac{T}{w({\bm p}_1)} \ .
\label{eq:2.18}
\ee
The absence of Landau interaction parameters in the noise correlation reflects the fact that the interactions included in this
model are purely deterministic and do not contribute to dissipative processes. See also Appendix~\ref{app:D}, where
we prove an H-theorem for the nonlinear Landau kinetic equation.

\bigskip

\section{Applications of the kinetic theory}
\label{sec:III}

\subsection{The equilibrium structure factor in the collisionless regime}
\label{subsec:III.A}

In this section we use the fluctuating kinetic equation derived in the previous section to calculate the dynamic
structure factor $S({\bm k},\omega)$ of a clean neutral Fermi liquid in the collisionless regime, which is always realized
at sufficiently low temperatures and sufficiently large wave numbers. In this limit, the streaming and interaction terms
in the kinetic operator $L_{\bm k}({\bm p})$ dominate over the collision operator, and $-\Lambda({\bm p})$ is replaced 
by a positive infinitesimal constant
$\epsilon$ that regularizes the free-fermion propagator. The quasiparticle interaction is described by the Landau
function $F({\bm p}_1,{\bm p}_2)$. In order to make contact with well-known results from conventional many-body
theory\cite{Fetter_Walecka_1971} we restrict the calculation to a model where the latter is a constant,
$F({\bm p}_1,{\bm p}_2) = F_0$. Equation~(\ref{eq:2.6f}) then implies ${\bm v}_p = {\bm p}/m$, and the kinetic
equation (\ref{eq:2.6b}) takes the form
\bea
{\hat\phi}({\bm p},{\bm k},\omega) &=& G_0({\bm p},{\bm k},\omega)\,{\hat F}_{\text{L}}({\bm p},{\bm k},\omega)
\nonumber\\
   && - \frac{{\bm p}\cdot{\bm k}}{m}\,G_0({\bm p},{\bm k},\omega)\,\frac{F_0}{N_0}\,\delta{\hat n}({\bm k},\omega)\ .
\label{eq:3.1}
\eea
Here
\be
G_0({\bm p},{\bm k},\omega) = \frac{i}{\omega + i\epsilon - {\bm p}\cdot{\bm k}/m}
\label{eq:3.2}
\ee
is the free-fermion Green function, and
\be
\delta{\hat n}({\bm k},\omega) = \frac{1}{V} \sum_{\bm p} \delta{\hat f}({\bm p},{\bm k},\omega) = \langle 1 \vert {\hat\phi}({\bm p},{\bm k},\omega)\rangle
\label{eq:3.3}
\ee
is the number-density fluctuation operator. 

The structure factor is given in terms of the density-density anticommutator correlation function via the relation\cite{correlation_functions_footnote}
\begin{widetext}
\bse
\label{eqs:3.4}
\be
\frac{1}{2} \big\langle \left[\delta{\hat n}({\bm k}_1,\omega_1),\delta{\hat n}({\bm k}_2,\omega_2)\right]_+ \big\rangle  = 
     2\pi \delta(\omega_1 + \omega_2)\,V \delta_{{\bm k}_1+{\bm k}_2,0}\,S({\bm k}_1,\omega_1)\ .
\label{eq:3.4a}
\ee
Via the fluctuation-dissipation theorem, Eq.~(\ref{eq:2.8}), it is related to the density-density response function
\be
\chi''_{nn}({\bm k},\omega) = \frac{1}{V^2}\sum_{{\bm p}_1,{\bm p}_2} \chi''({\bm p}_1,{\bm p}_2;{\bm k},\omega) 
     = \tanh(\omega/2T) S({\bm k},\omega)\ .
\label{eq:3.4b}
\ee
\ese
From Eqs.~(\ref{eq:3.1}) and (\ref{eq:3.3}) we obtain an explicit expression for $\delta{\hat n}({\bm k},\omega)$,
\bse
\label{eqs:3.5}
\be
\delta{\hat n}({\bm k},\omega) = \frac{1}{1 + F_0 J_1({\bm k},\omega)}\,\frac{1}{V}\sum_{\bm p} w({\bm p})\,G_0({\bm p},{\bm k},\omega)\,{\hat F}_{\text{L}}({\bm p},{\bm k},\omega)\ ,
\label{eq:3.5a}
\ee
where
\bea
J_1({\bm k},\omega) &=& \frac{-1}{N_0V} \sum_{\bm p} w({\bm p})\,\frac{{\bm p}\cdot{\bm k}/m}{\omega + i\epsilon - {\bm p}\cdot{\bm k}/m}
\nonumber\\
&=& J_1(-{\bm k},-\omega)^*\ .
\label{eq:3.5b}
\eea
\ese
By using Eq.~(\ref{eq:3.5a}) in (\ref{eq:3.4a}) we see that the structure factor can be expressed in terms of the symmetrized
Langevin force correlation $\Psi_+$. Using Eq.~(\ref{eq:2.11}) we find
\be
S({\bm k},\omega) =  \frac{1}{1 + F_0 J_1({\bm k},\omega)}\, \frac{1}{1 + F_0 J_1(-{\bm k},-\omega)}\,
\frac{1}{V^2}\sum_{{\bm p}_1,{\bm p}_2} w({\bm p}_1)\,w({\bm p}_2)\,G_0({\bm p}_1,{\bm k},\omega)\,G_0({\bm p}_2,-{\bm k},-\omega)\,
     \Psi_+({\bm p}_1,{\bm p}_2;{\bm k},\omega)\ .
\label{eq:3.6}
\ee    
 We now perform the zero-temperature limit in the terms where we have not done so yet. Then we have
 \bse
 \label{eqs:3.7}
 \bea
 -G_0({\bm p}_1,{\bm k},\omega)\,G_0({\bm p}_2,-{\bm k},-\omega) \left[\Lambda({\bm p}_1) + \Lambda({\bm p}_2\right]\delta_{{\bm p}_1,{\bm p}_2}   
      &\to&  \delta_{{\bm p}_1,{\bm p}_2} 2\pi \delta(\omega - {\bm p}_1\cdot{\bm k}/m)
\label{eq:3.7a}\\
\omega\coth(\omega/2T) &\to& \omega\, \sgn\omega
\label{eq:3.7b}
\eea
\ese
\end{widetext}
Finally, the imaginary part of $J_1$ is
\be
\Im\,J_1({\bm k},\omega) = \frac{\pi}{N_0V}\sum_{\bm p} w({\bm p})\,\frac{{\bm p}\cdot{\bm k}}{m}\,\delta(\omega - {\bm p}\cdot{\bm k}/m)\ .
\label{eq:3.8}
\ee
Combining all of this we obtain
\be
S({\bm k},\omega) = N_0\,\frac{\sgn\omega\,\Im J_1({\bm k},\omega)}{\left(1 + F_0\,\Re J_1({\bm k},\omega)\right)^2 + \left(F_0\,  \Im\,J_1({\bm k},\omega)\right)^2}
\label{eq:3.9}
\ee   
We recognize this as the standard result obtained from many-body theory in the so-called ring or random-phase approximation.\cite{Fetter_Walecka_1971}
See also Ref.~\onlinecite{Belitz_Kirkpatrick_2012a}, where the same result was obtained by field-theoretic methods. Its physical contents include the well known
$\ell=0$ zero-sound excitation and the particle-hole continuum that were discussed in Paper I.

\subsection{Fluctuating Navier-Stokes equations for a fermionic quantum fluid}
\label{subsec:III.B}

We now consider the hydrodynamic regime, where the collision operator dominates over the operator $L_{\bm k}^{(1)}$. 
It  is realized in the limit of small frequencies, $\omega\tau\ll 1$, and wave numbers, $k\ell\ll 1$, with $\tau$ a
generic collision time and $\ell$ a generic mean-free path. As in the previous subsection, we consider a neutral Fermi liquid; 
at the end of this subsection we will discuss the modifications necessary for discussing metals. For simplicity we will use
the simple quasiparticle model interaction given in Eq.~(\ref{eq:2.16a}). Within this model, ${\bm v}_p = {\bm p}/m^*$ and
$\epsilon_p = {\bm p}^2/2m^*$. We ignore a $p$-independent contribution to $\epsilon_p$ that also depends on the
FL interaction,\cite{Fetter_Walecka_1971} as we did in Paper I.

The quantum mechanical generalization of the Boltzmann equation for this case was first formulated by Uehling and 
Uhlenbeck.\cite{Uehling_Uhlenbeck_1933} A crucial aspect of any kinetic theory for a fluid, quantum or classical, is that 
the linearized collision operator has five (or $d+2$ in $d$ spatial dimensions) zero eigenvalues, and corresponding zero 
eigenfunctions, that represent the conservation of particle number, momentum, and energy.

The relevant hydrodynamic variables are the density fluctuation $\delta{\hat n}$, the three components of the fluid velocity
fluctuation $\delta{\hat u}_i$ ($i=x,y,z$), and the temperature fluctuation $\delta{\hat T}$. With the scalar product defined
in Eq.~(\ref{eq:2.3a}) they can be expressed as follows:
\bse
\label{eqs:3.10}
\bea
\delta{\hat n}({\bm k},\omega) &=& \langle a_1({\bm p}) \vert \hat{\bm\phi}({\bm p},{\bm k},\omega)\rangle\ ,
\label{eq:3.10a}\\
\delta{\hat u}_i ({\bm k},\omega) &=& \frac{1}{nm} \langle a_i({\bm p}) \vert \hat{\bm\phi}({\bm p},{\bm k},\omega)\rangle\ ,
\label{eq:3.10b}\\
\delta{\hat T}({\bm k},\omega) &=& \frac{1}{c_V}\,\langle a_5({\bm p}) \vert \hat{\bm\phi}({\bm p},{\bm k},\omega)\rangle\ ,
\label{eq:3.10c}
\eea
where $c_V$ is the specific heat at constant volume. Also of interest is the entropy density fluctuation
\be
\delta{\hat s} ({\bm k},\omega) = \frac{1}{T} \left\langle\epsilon_p - \mu \Big\vert {\hat\phi}({\bm p},{\bm k},\omega)\right\rangle\ ,
\label{eq:3.10d}
\ee
the pressure fluctuation
\be
\delta{\hat p}({\bm k},\omega) = \left(\frac{\partial p}{\partial n}\right)_{T,V} \delta{\hat n}({\bm k},\omega) +  \left(\frac{\partial p}{\partial T}\right)_{N,V} \delta{\hat T}({\bm k},\omega) \ ,
\label{eq:3.10e}
\ee
and the energy density fluctuation
\be
\delta{\hat e}({\bm k},\omega) = \langle \epsilon_p \vert {\hat\phi}({\bm p},{\bm k},\omega)\rangle\ ,
\label{eq:3.10f}
\ee
\ese
The five functions $a_{\alpha}({\bm p})$ are
\bse
\label{eqs:3.11}
\bea
a_1({\bm p}) &=& 1\ ,
\label{eq:3.11a}\\
a_i({\bm p}) &=& p_i\qquad (i=x,y,z)\ ,
\label{eq:3.11b}\\
a_5({\bm p}) &=& \epsilon_p - \langle\epsilon_p\rangle_w\ .
\label{eq:3.11c}
\eea
\ese
The validity of Eqs.~(\ref{eq:3.10a}, \ref{eq:3.10b},  \ref{eq:3.10e},  \ref{eq:3.10f}) is obvious.
For a derivation of Eq.~(\ref{eq:3.10c}) see Appendix A 2 in Paper I,
and for (\ref{eq:3.10d}) see Eq.~(3.18) in Paper I.

The five functions $a_{\alpha}({\bm p})$ are eigenfunctions of the collision operator for the five zero eigenvalues that
represent the conservation of particle number, momentum, and energy,
\be
\Lambda({\bm p})\,\vert a_{\alpha}({\bm p})\rangle = 0\ .
\label{eq:3.12}
\ee
They span the hydrodynamic subspace ${\cal L}_0$. This, combined 
with the fact that the collisions dominate the physics, implies that one can derive hydrodynamic equations by using a projection 
operator ${\cal P}$ defined by
\bse
\label{eqs:3.13}
\be
{\cal P} = \sum_{\alpha} \frac{\vert a_{\alpha}({\bm p})\rangle\langle a_{\alpha}({\bm p})\vert}{\langle  a_{\alpha}({\bm p}) \vert  a_{\alpha}({\bm p})\rangle}\ .
\label{eq:3.13a}
\ee
that projects onto ${\cal L}_0$. The projector onto the complementary space ${\cal L}_{\perp}$ is
\be
{\cal P}_{\perp} = \openone - {\cal P}\ ,
\label{eq:3.13b}
\ee
\ese
where $\openone$ is the unit operator. 

\subsubsection{The projected $\mu$-space distribution}
\label{subsubsec:III.B.1}

Now consider the kinetic equation (\ref{eq:2.6b}) and operate from the left with ${\cal P}$. 
As in the classical Boltzmann-Langevin equation,\cite{Dorfman_vanBeijeren_Kirkpatrick_2021} the fluctuating force has no overlap with the five
zero eigenfunctions $a_{\alpha}({\bm p})$ of the collision operator, Eqs.~(\ref{eqs:3.11}, \ref{eq:3.12}). This can be seen explicitly from Eq.~(\ref{eq:2.18}):
The function $\Psi_-$ has no overlap with either $a_{\alpha}({\bm p}_1)$ or $a_{\alpha}({\bm p}_2)$. Since the (zero) first moment of ${\hat F}_{\text{L}}$
and the second moment $\Psi_-$ completely determine the distribution we thus have
\bse
\label{eqs:3.14}
\be
{\cal P} \vert{\hat F}_{\text{L}}({\bm p},{\bm k},\omega)\rangle = 0\ .
\label{eq:3.14a}
\ee
Equation~(\ref{eq:2.6b}) then implies
\be
\left[ -i\omega {\cal P} + {\cal P} L_{\bm k}^{(1)}({\bm p})\right] \vert{\hat\phi}({\bm p},{\bm k},\omega)\rangle = 0\ .
\label{eq:3.14b}
\ee
\ese
By using Eq.~(\ref{eq:3.13b}) this can be written
\begin{widetext}
\be
\left[-i\omega + {\cal P}  L_{\bm k}^{(1)}({\bm p})\right] {\cal P} \vert {\hat\phi}({\bm p},{\bm k},\omega)\rangle 
   = -{\cal P}  L_{\bm k}^{(1)}({\bm p}) {\cal P}_{\perp} \vert{\hat\phi}({\bm p},{\bm k},\omega)\rangle 
\label{eq:3.15}
\ee
Our goal is to obtain a closed equation for the projected $\mu$-space distribution ${\cal P}\vert{\hat\phi}\rangle$. To this
end we multiply Eq.~(\ref{eq:2.6b}) from the left with ${\cal P}_{\perp}$, which yields
\be
\left[-i\omega + {\cal P}_{\perp}  L_{\bm k}^{(1)}({\bm p}) - \Lambda({\bm p})\right] {\cal P}_{\perp} \vert{\hat\phi}({\bm p},{\bm k},\omega) \rangle
     = -{\cal P}_{\perp}  L_{\bm k}^{(1)}({\bm p}) {\cal P}\vert{\hat\phi}({\bm p},{\bm k},\omega)\rangle + {\cal P}_{\perp} \vert{\hat F}_{\text L}({\bm p},{\bm k},\omega)\rangle\ .
\label{eq:3.16}
\ee
Here we have used Eq.~(\ref{eq:3.12}), which implies ${\cal P}_{\perp}\Lambda({\bm p}) = \Lambda({\bm p}){\cal P}_{\perp} = \Lambda({\bm p})$. 
 
We are interested in hydrodynamic equations that are valid for small wave numbers up to $O(k^2)$, with the fluctuating force included to lowest
order in $k$. In a neutral Fermi liquid the hydrodynamic frequency scales either linearly or quadratically with $k$, while the collision operator acting on the orthogonal 
space ${\cal L}_{\perp}$ scales as a constant.\cite{collision_operator_footnote} Therefore, the first two terms in brackets on the left-hand side of 
Eq.~(\ref{eq:3.16}) are small corrections to $\Lambda$ and we can approximate
${\cal P}_{\perp}{\hat\phi}$ by
\be
{\cal P}_{\perp} \vert{\hat\phi}({\bm p},{\bm k},\omega)\rangle \approx \Lambda_{\perp}^{-1}({\bm p}) 
      L_{\bm k}^{(1)}({\bm p}) {\cal P} \vert{\hat\phi}({\bm p},{\bm k},\omega)\rangle   
 - \Lambda_{\perp}^{-1}({\bm p}) \vert {\hat F}_{\text L}({\bm p},{\bm k},\omega)\rangle\ .
\label{eq:3.17}
\ee   
Here $\Lambda_{\perp}^{-1}({\bm p}) =  {\cal P}_{\perp} \left(\Lambda({\bm p})\right)^{-1} {\cal P}_{\perp}$, and we have used
the projector property ${\cal P}_{\perp}^2 = {\cal P}_{\perp}$.
Corrections are of relative $O(k)$. By using Eq.~(\ref{eq:3.17}) in (\ref{eq:3.15}) we obtain a closed formal equation for ${\cal P}{\hat\phi}$,
\be
\left[-i\omega + {\cal P}  L_{\bm k}^{(1)}({\bm p})\right] {\cal P} \vert{\hat\phi}({\bm p},{\bm k},\omega)\rangle = - {\cal P}  L_{\bm k}^{(1)}({\bm p}) 
     \Lambda_{\perp}^{-1}({\bm p}) L_{\bm k}^{(1)}({\bm p}) {\cal P} \vert{\hat\phi}({\bm p},{\bm k},\omega)\rangle
          + {\cal P} L_{\bm k}^{(1)}({\bm p}) \Lambda_{\perp}^{-1}({\bm p}) \vert{\hat F}_{\text L}({\bm p},{\bm k},\omega)\rangle\ .
\label{eq:3.18}
\ee
\end{widetext}

\subsubsection{The fluctuating Navier-Stokes equations}
\label{subsubsec:III.B.2}

We can now derive the five hydrodynamic equations by multiplying Eq.~(\ref{eq:3.18}) from the left with $\langle a_{\alpha}({\bm p})\vert$. 
For the remainder of this subsection we will write the hydrodynamic quantities as functions of the real-space position ${\bm x}$ rather
than the wave vector ${\bm k}$.

\paragraph{Density equation}
\label{par:III.B.2.a}

Multiplying Eq.~(\ref{eq:3.18}) from the left with $\langle 1\vert$ yields the continuity equation for the density,
\be
-i\omega\,\delta{\hat n}({\bm x},\omega) + n{\bm\nabla}\cdot \delta{\hat{\bm u}}({\bm x},\omega) = 0\ .
\label{eq:3.19}
\ee
The number current density given by $n{\hat{\bm u}}$ is exact, and neither the collision operator nor the
fluctuating force enter the density equation. 

\paragraph{Fluid velocity equation}
\label{par:III.B.2.b}

The equation for the velocity, which we obtain by multiplying Eq.~(\ref{eq:3.18}) from the left with $\langle{\bm p}\vert$, is more involved. 
The projector $P_{\perp}$ acting on functions that are even (odd) in ${\bm p}$
yields functions that also are even (odd) in ${\bm p}$, and the inverse collision operator $\Lambda^{-1}({\bm p})$ is isotropic
in momentum space. As a result, the angular integrations make sure that the contributions from the various modes on the
right-hand side of Eq.~(\ref{eq:3.18}) do not couple, and we find
\bse
\label{eqs:3.20}
\be
-i\omega\, nm\, \delta{\hat u}_i({\bm x},\omega) + \partial_j\, {\hat\tau}_{ij}({\bm x},\omega) + \partial_j \left({\hat\tau}_{\text{L}}\right)_{ij}({\bm x},\omega) = 0\ .
\label{eq:3.20a}
\ee
Here 
\bea
{\hat\tau}_{ij}({\bm x},\omega) &=& \delta_{ij}\, \delta{\hat p}({\bm x},\omega) 
\nonumber\\
&& \hskip -50pt - \eta \left[\partial_i\, \delta{\hat u}_j({\bm x},\omega) + \partial_j\, {\hat u}_i({\bm x},\omega) - \frac{2}{3}\,\delta_{ij}\, {\bm\nabla}\cdot{\hat{\bm u}}({\bm x},\omega) \right] 
\nonumber\\
&& + \zeta\,\delta_{ij}\,{\bm\nabla}\cdot{\hat{\bm u}}({\bm x},\omega)
\label{eq:3.20b}
\eea
is the stress tensor. Its reactive part is given in terms of the pressure fluctuation $\delta{\hat p}$, Eq.~(\ref{eq:3.10e}).
The dissipative part has the form familiar from the classical Navier-Stokes equations.\cite{Forster_1975, Landau_Lifshitz_VI_1959}
$\eta$ is the shear viscosity given by
\be
\eta = - {\hat k}_{\perp}^i {\hat k}^j \,{\hat k}_{\perp}^l {\hat k}^m \left\langle \sigma_{ij}({\bm p})\vert\Lambda^{-1}({\bm p})\vert \sigma_{lm}({\bm p})\right\rangle\ .
\label{eq:3.20c}
\ee
Here $\hat{\bm k}$ is the unit vector in the direction of ${\bm k}$, and $\hat{\bm k}_{\perp}$ is either of the two 
vectors perpendicular to ${\bm k}$. The $\mu$-space momentum current $\sigma_{ij}$ is
\be
\vert\sigma^{ij}({\bm p})\rangle = \vert\sigma_1^{ij}({\bm p})\rangle - \delta^{ij}\,\vert\sigma_2({\bm p})\rangle
\label{eq:3.20d}
\ee
with
\be
\vert\sigma_1^{ij}({\bm p})\rangle = \vert p^i v_p^j\rangle - \delta^{ij} \frac{1}{3}\,\vert{\bm p}\cdot{\bm v}_p\rangle\ .
\label{eq:3.20e}
\ee
and
\be
\vert\sigma_2({\bm p})\rangle = \frac{-1}{3}\,{\cal P}_{\perp} \vert{\bm p}\cdot{\bm v}_p\rangle\ .
\label{eq:3.20f}
\ee
$\zeta$ is a contribution to the bulk viscosity given by
\be
\zeta = - \langle\sigma_2({\bm p})\vert\Lambda^{-1}({\bm p})\vert\sigma_2({\bm p})\rangle\ .
\label{eq:3.20g}
\ee
Note that within the simple model characterized by Eq.~(\ref{eq:2.16a}), ${\bm p}\cdot{\bm v}_p = 2\epsilon_p \in {\cal L}_0$,
and hence $\sigma_2({\bm p})$ vanishes, but it is nonzero
in general. For an expression of $\sigma_2$ in terms of thermodynamic quantities see Paper I. Finally, 
\be
\left({\hat\tau}_{\text{L}}\right)_{ij}({\bm x},\omega) = \left\langle\sigma_{ij}({\bm p})\vert\Lambda^{-1}({\bm p})\vert {\hat F}_{\text{L}}({\bm p},{\bm x},\omega)\right\rangle
\label{eq:3.20h}
\ee
\ese
is the fluctuating part of the stress tensor. In deriving Eqs.~(\ref{eqs:3.20}) we have made use of various thermodynamic relations that
are derived in Appendix A of Paper I; the most important ones are also given in Appendix~\ref{app:E}.

\smallskip
\paragraph{Temperature or heat equation}
\label{par:III.B.2.c}

Finally, multiplying Eq.~(\ref{eq:2.18}) from the left with $\langle a_5({\bm p})\vert$ yields the heat equation. The calculation yields
\bse
\label{eqs:3.21}
\bea
-i\omega\,c_V \delta{\hat T}({\bm x},\omega) &+& T\left(\frac{\partial p}{\partial T}\right)_{N,V} {\bm\nabla}\cdot{\hat{\bm u}}({\bm x},\omega) 
\nonumber\\
&&\hskip -60pt - \kappa\,{\bm\nabla}^2 \delta{\hat T}({\bm x},\omega) + {\bm\nabla}\cdot{\hat{\bm q}}_{\text{L}}({\bm x},\omega)= 0\ .
\label{eq:3.21a}
\eea
Here 
\be
\kappa = \frac{-1}{T}\,{\hat k}_i {\hat k}_j  \left\langle v_p^i\,\psi_5^{\text{L}(0)}({\bm p})\vert\Lambda^{-1}({\bm p})\vert v_p^j\,\psi_5^{\text{L}(0)}({\bm p})\right\rangle
\label{eq:3.21b}
\ee
is the heat conductivity with
\be
\psi_5^{\text{L}(0)}({\bm p}) = a_5({\bm p}) - \frac{T}{n}\left(\frac{\partial p}{\partial T}\right)_{N,V} a_1({\bm p})
\label{eq:3.21c}
\ee
the heat mode to zeroth order in the wave vector ${\bm k}$; see Paper I for the identification of the heat mode and Appendix~\ref{app:C} for
a relation between the present approach and the kinetic-theory approach in Paper I. The final term in Eq.~(\ref{eq:3.21a}) is the
fluctuating heat current, given by
\be
{\hat{\bm q}}_{\text{L}}({\bm x},\omega) = - \langle {\bm v}_p\, \psi_5^{\text{L}(0)}({\bm p})\vert\Lambda^{-1}({\bm p})\vert {\hat F}_{\text{L}}({\bm p},{\bm x},\omega)\rangle\ .
\label{eq:3.21d}
\ee
\ese

Alternatively, we can write the heat equation in terms of the time derivative of either the energy density fluctuation 
$\delta{\hat e}$ or the entropy density fluctuation $\delta{\hat s}$  instead of $\delta{\hat T}$. 
From Eqs.~(\ref{eq:3.21a}) and (\ref{eq:3.19}), and using the identity (\ref{eq:E.2}), we find the heat equation in the form\cite{Forster_1975}
\bea
-i\omega\,\delta {\hat e}({\bm x},\omega) &+& (e+p){\bm\nabla}\cdot \delta{\hat{\bm u}}({\bm x},\omega) - \kappa\,{\bm\nabla}^2 \delta{\hat T}({\bm x},\omega) 
\nonumber\\
&+& {\bm\nabla}\cdot{\hat{\bm q}}_{\text{L}}({\bm x},\omega)= 0\ .
\label{eq:3.22}
\eea   
Finally, using Eq.~(\ref{eq:3.19}) to eliminate the ${\bm\nabla}\cdot\delta{\hat{\bm u}}$ term in (\ref{eq:3.22}) yields it in the 
form\cite{Landau_Lifshitz_VI_1959, Landau_Lifshitz_IX_1991}
\bse
\label{eqs:3.23}
\be
 -i\omega\,Tn\,\delta {\hat s}_n({\bm x},\omega)  - \kappa\,{\bm\nabla}^2 \delta{\hat T}({\bm x},\omega) + {\bm\nabla}\cdot{\hat{\bm q}}_{\text{L}}({\bm x},\omega)= 0\ ,
\label{eq:3.23a}
\ee
where 
\be
\delta {\hat s}_n = \frac{1}{n}\,\delta {\hat s} - \frac{s}{n^2}\, \delta {\hat n} = \frac{1}{Tn} \left(\delta{\hat e} - \frac{e+p}{n}\, \delta{\hat n}\right)
\label{eq:3.32b}
\ee
\ese
is the fluctuation of the entropy per particle (see Eq.~(3.18) in Paper I).
Equations (\ref{eq:3.21a}), (\ref{eq:3.22}), and (\ref{eqs:3.23}) are pairwise equivalent.
    
\smallskip
\paragraph{Fluctuating force correlations}
\label{par:III.B.2.d}
     
We still need to determine the correlations of the fluctuating forces in the hydrodynamic equations. From Eqs.~(\ref{eq:2.11}),
(\ref{eq:2.15c}), and (\ref{eq:2.18}), and repeatedly using the fact that $\Lambda({\bm p})$ and $\Lambda^{-1}({\bm p})$ are
self-adjoint with respect to the scalar product $\langle\ldots\vert\ldots\rangle$,  we find for the anticommutator correlation of 
the fluctuating stress tensor, Eq.~(\ref{eq:3.20h}),
\begin{widetext}
\bse
\label{eqs:3.24}
\be
\frac{1}{2} \Big\langle  \left[\left({\hat\tau}_{\text{L}}\right)_{ij}({\bm x}_1,\omega_1),\left({\hat\tau}_{\text{L}}\right)_{kl}({\bm x}_2,\omega_2)\right]_+ \Big\rangle  = 2\pi \delta(\omega_1+\omega_2)\,\delta({\bm x}_1 - {\bm x}_2)\,\omega_1 \coth(\omega_1/2T) \left[\eta (\delta_{ik} \delta_{jl} + \delta_{il} \delta_{jk})
   + (\zeta - \frac{1}{3}\,\eta) \delta_{ij} \delta_{kl} \right]\ .
\label{eq:3.24a}
\ee
Similarly, the anticommutator correlation of the fluctuating heat current, Eq.~(\ref{eq:3.21d}), is
\be
\frac{1}{2} \big\langle  [q_{\text{L}}^i({\bm x}_1,\omega_1),q_{\text{L}}^j({\bm x}_2,\omega_2)]_+ \big\rangle  =  \delta^{ij}
     2\pi \delta(\omega_1+\omega_2)\,\delta({\bm x}_1 - {\bm x}_2)\,\omega_1 T \coth(\omega_1/2T)\,\kappa\ .
\label{eq:3.24b}
\ee
and the cross correlations vanish,
\be
\frac{1}{2} \Big\langle  \left[\left({\hat\tau}_{\text{L}}\right)_{ij}({\bm x}_1,\omega_1), q_{\text{L}}^k({\bm x}_2,\omega_2)\right]_+ \Big\rangle  = 0\ .
\label{eq:3.24c}
\ee
\ese 
\end{widetext}   
The corresponding commutator correlations are given by the same expressions with $\coth(\omega_1/2T)$ replaced by $1$.

The equations (\ref{eqs:3.24}) are identical to those given in \S 88 of Ref.~\onlinecite{Landau_Lifshitz_IX_1991} without a derivation.

\subsection{Fluctuating hydrodynamic equations for a disordered Fermi liquid}
\label{subsec:III.C}

We now consider fermions in the presence of quenched disorder. In this situation particle number and energy are still
conserved, but the fermion momentum is not. For a derivation of hydrodynamic equations for such a system we can
still use Eq.~(\ref{eq:3.18}), but the projector ${\cal P}$ now projects on a space that is spanned by $a_1({\bm p})$ and
$a_5({\bm p})$ only. Accordingly, ${\cal P}\vert{\bm p}\rangle = {\cal P}\vert{\bm v}_p\rangle = 0$ and $\Lambda({\bm p})$
has only two zero eigenvalues with eigenvectors $a_1({\bm p})$ and $a_5({\bm p})$. In addition to the fermion-impurity
scattering $\Lambda({\bm p})$ can contain parts that describe other scattering processes, in particular fermion-fermion
scattering, as long as they conserve energy. In this subsection we use a model where only the LFL
parameter $F_0$ is nonzero. 

\subsubsection{Fluctuating hydrodynamic equations}
\label{subsubsec:III.C.1}

\paragraph{Density equation}
\label{par:III.C.1.a}

Multiplying Eq.~(\ref{eq:3.18}) from the left with $\langle a_1({\bm p})\vert$ yields an equation for the density fluctuation.
The streaming/interaction term on the left-hand side does not contribute, but the dissipative and fluctuating terms both
do. A calculation that is analogous to the one in Sec.~\ref{subsubsec:III.B.2} yields
\bse
\label{eqs:3.25}
\bea
-i \omega\,\delta{\hat n}({\bm x},\omega) &=& D_{11} {\bm\nabla}^2\,\delta{\hat n}({\bm x},\omega) + D_{12}\, {\bm\nabla}^2\,\delta{\hat T}({\bm x},\omega) 
\nonumber\\
&&\hskip 50pt - {\bm\nabla}\cdot {\bm j}_{\text{L}}({\bm x},\omega)\ .
\label{eq:3.25a}
\eea
Here
\be
D_{11} = \frac{-1}{m^2}\,\frac{1}{(\partial n/\partial\mu)_{T,V}}\,\langle{\hat{\bm k}}\cdot{\bm p}\vert\Lambda^{-1}({\bm p})\vert{\hat{\bm k}}\cdot{\bm p}\rangle\ ,
\label{eq:3.25b}
\ee
and
\be
D_{12} =  \frac{-1}{m^2}\,\frac{1}{T}\,\langle{\hat{\bm k}}\cdot{\bm p}\vert\Lambda^{-1}({\bm p})\vert ({\hat{\bm k}}\cdot{\bm p})a_5({\bm p})\rangle
\label{eq:3.25c}
\ee
are transport coefficients and
\be
{\bm j}_{\text{L}}({\bm x},\omega)  = -\,\langle{\bm v}_p\vert\Lambda^{-1}({\bm p})\vert{\hat F}({\bm p},{\bm x},\omega)\rangle
\label{eq:3.25d}
\ee
\ese
is a fluctuating force that dimensionally is a number current density. Note that the absence of momentum conservation completely changes
the structure of the density equation compared to the continuity equation in Sec.~\ref{par:III.B.2.a}.

\smallskip
\paragraph{Heat equation}
\label{par:III.C.1.b}

Multiplying Eq.~(\ref{eq:3.18}) from the left with $\langle a_5({\bm p})\vert$ yields 
\bse
\label{eqs:3.26}
\bea
-i\omega\,c_V \delta{\hat T}({\bm x},\omega) &=& \frac{T}{(\partial n/\partial\mu)_{N,V}}\,D_{12}\,{\bm\nabla}^2 \delta{\hat n}({\bm x},\omega)
\nonumber\\
&&\hskip -50pt + {\bar\kappa}\,{\bm\nabla}^2 \delta{\hat T}({\bm x},\omega) - {\bm\nabla}\cdot{\bm{\bar q}}({\bm x},\omega)\ .
\label{eq:3.26a}
\eea
Here
\be
{\bar\kappa} = \frac{-1}{T}\, {\hat k}_i {\hat k}_j  \left\langle v_p^i\,a_5({\bm p})\vert\Lambda^{-1}({\bm p})\vert v_p^j\,a_5({\bm p})\right\rangle
\label{eq:3.26b}
\ee
is a modified heat conductivity, and
\be
{\bm{\bar q}}_{\text{L}}({\bm x},\omega)  = - \langle a_5({\bm p}) {\bm v}_p\vert\Lambda^{-1}({\bm p})\vert{\hat F}({\bm p},{\bm x},\omega)\rangle
\label{eq:3.26c}
\ee
\ese
is the relevant fluctuating force.

\smallskip
\paragraph{Fluctuating force correlations}
\label{par:III.C.1.c}

For the anticommutator correlations of the fluctuating forces we find
\begin{widetext}
\bse
\label{eqs:3.27}
\bea
\frac{1}{2} \big\langle  [v_{\text{L}}^i({\bm x}_1,\omega_1),v_{\text{L}}^j({\bm x}_2,\omega_2)]_+ \big\rangle  &=&  \delta^{ij}
     2\pi \delta(\omega_1+\omega_2)\,\delta({\bm x}_1 - {\bm x}_2)\,\omega_1 \coth(\omega_1/2T)\,\frac{1}{T n^2}\,D_{11}\ ,
\label{eq:3.27a}\\
\frac{1}{2} \big\langle  [{\bar q}_{\text{L}}^i({\bm x}_1,\omega_1),{\bar q}_{\text{L}}^j({\bm x}_2,\omega_2)]_+ \big\rangle  &=& \delta^{ij}
     2\pi \delta(\omega_1+\omega_2)\,\delta({\bm x}_1 - {\bm x}_2)\,\omega_1 \coth(\omega_1/2T)\,{\bar\kappa}\ ,
\label{eq:3.27b}\\
\frac{1}{2} \big\langle  [{\bar q}_{\text{L}}^i({\bm x}_1,\omega_1),v_{\text{L}}^j({\bm x}_2,\omega_2)]_+ \big\rangle  &=& \delta^{ij}
2\pi \delta(\omega_1+\omega_2)\,\delta({\bm x}_1 - {\bm x}_2)\,\omega_1 \coth(\omega_1/2T)\,\frac{1}{n}\,D_{12}\ .
\eea
\label{eq:3.27c}
\ese
\end{widetext}
The corresponding commutator correlations are again given by the same expressions with $\coth(\omega_1/2T)$ replaced by $1$.

\subsubsection{Onsager relations}
\label{subsucsec:III.C.2}

We now show that the results in Sec.~\ref{subsubsec:III.C.1} are consistent with irreversible thermodynamics and
the Onsager reciprocal relations. 

In the absence of momentum conservation the relevant currents are the number current density ${\bm j}_n$ and the heat or entropy current density 
${\bm j}_s$.\cite{Callen_1985, Mahan_1981} They are driven by gradients of the chemical potential and the temperature, which defines Onsager 
coefficients $L_{ij}$ via the linear-response relations
\bse
\label{eqs:3.28}
\bea
{\bm j}_n &=& \frac{-L_{11}}{T}\,{\bm\nabla}\mu - \frac{L_{12}}{T^2}\,{\bm\nabla}T \ ,
\label{eq:3.28a}\\
{\bm j}_s &=& \frac{-L_{21}}{T}\,{\bm\nabla}\mu - \frac{L_{22}}{T^2}\,{\bm\nabla}T \ .
\label{eq:3.28b}
\eea
\ese
The relevant Onsager relation consists of the statement $L_{21} = L_{12}$.
The dynamic equations associated with these currents are
\bse
\label{eqs:3.29}
\bea
\partial_t n &=& -{\bm\nabla}\cdot{\bm j}_n\ ,
\label{eq:3.29a}\\
T\,\partial_t s &=& -{\bm\nabla}\cdot{\bm j}_s\ .
\label{eq:3.29b}
\eea
\ese
Equation~(\ref{eq:3.29a}) is the continuity equation for the number density, see Eq.~(B1a) in Paper I. Equation~(\ref{eq:3.29b})
expresses energy conservation. In order to relate it to Eq.~(B4a) in Paper I we note that the heat mode $\psi_5^{(0)}$ in that
paper is a linear combination of the entropy density and the number density fluctuations.\cite{entropy_footnote}

Ignoring fluctuations, the hydrodynamic equation (\ref{eq:3.25a}) yields
\be
{\bm j}_n = -D_{11} {\bm\nabla}n - D_{12} {\bm\nabla T}\ .
\label{eq:3.30}
\ee
In order to relate this to Eq.~(\ref{eq:3.28a}) we express ${\nabla\mu}$ in terms of ${\bm\nabla}n$ and ${\bm\nabla} T$ via
\bea
{\bm\nabla}\mu &=& \left(\frac{\partial\mu}{\partial n}\right)_{T,V} {\bm\nabla} n + \left(\frac{\partial\mu}{\partial T}\right)_{N,V} {\bm\nabla} T
\nonumber\\
&=& \frac{1}{(\partial n/\partial\mu)_{T,V}} {\bm\nabla}n - \langle\epsilon_p - \mu\rangle_w \frac{1}{T}\,{\bm\nabla}T\ ,\quad
\label{eq:3.31}
\eea
where we have used Eq.~(\ref{eq:E.3}). Comparing coefficients in Eqs.~(\ref{eq:3.28a}) and (\ref{eq:3.30}) yields
\bea
L_{11} &=& \frac{-T}{m^2}\,\langle{\hat{\bm k}}\cdot{\bm p}\vert\Lambda^{-1}({\bm p})\vert{\hat{\bm k}}\cdot{\bm p}\rangle\ ,
\label{eq:3.32}\\
L_{12} &=&  \frac{-T}{m^2}\,\langle{\hat{\bm k}}\cdot{\bm p}\vert\Lambda^{-1}({\bm p})\vert ({\hat{\bm k}}\cdot{\bm p})(\epsilon_p - \mu)\rangle\ .
\label{eq:3.33}
\eea

In order to obtain the remaining two Onsager coefficients we note that entropy density, temperature, and number density fluctuations
are related by (see Eq.~(3.18) in Paper I)
\bse
\label{eqs:3.34}
\be
T \delta s = c_V \delta T - T(\partial\mu/\partial T)_{N,V} \delta n\ .
\label{eq:3.34a}
\ee
With Eqs.~(\ref{eqs:3.29}) this yields
\be
c_V \partial_t T = -{\bm\nabla}\cdot{\bm j}_s - T \left(\frac{\partial\mu}{\partial T}\right)_{N,V} {\bm\nabla}\cdot{\bm j}_n\ .
\label{eq:3.34b}
\ee
\ese
By using Eqs.~(\ref{eqs:3.26}), neglecting the fluctuating force term, this allows us to express ${\bm j}_s$ in terms of
density and temperature gradients,
\bea
{\bm j}_s &=& \left[T\left(\frac{\partial\mu}{\partial T}\right)_{N,V} D_{11} - T \left(\frac{\partial\mu}{\partial n}\right)_{T.V} D_{12}\right] {\bm\nabla} n
\nonumber\\
&& + \left[ T\left(\frac{\partial\mu}{\partial T}\right)_{N,V} D_{12} - {\bar\kappa}\right] {\bm\nabla} T\ .
\label{eq:3.35}
\eea
By comparing the coefficients in Eqs.~(\ref{eq:3.35}) and (\ref{eq:3.28b}), and using Eqs.~(\ref{eq:3.31}) and (\ref{eq:E.3}) we find
\bea
L_{21} &=& L_{12}\ ,
\label{eq:3.36}\\
L_{22} &=& \frac{-T}{m^2} \langle ({\hat{\bm k}}\cdot{\bm p})(\epsilon_p - \mu)\vert\Lambda^{-1}({\bm p})\vert ({\hat{\bm k}}\cdot{\bm p})(\epsilon_p - \mu)\rangle\ .
\nonumber\\
\label{eq:3.37}
\eea
Equation~(\ref{eq:3.36}) is the required Onsager reciprocal relation. Note that the heat current that appears in Eqs.~(\ref{eq:3.33}) and
(\ref{eq:3.37}) is the one associated with the entropy density, Eq.~(\ref{eq:3.10d}). It is different from the heat current that determines
the thermal conductivity in a fluid with momentum conservation (which is associated with the entropy per particle, see Paper I), and
also from the one that determines the transport coefficient $\bar\kappa$, Eq.~(\ref{eq:3.26b}).

\section{Discussion, and Outlook}
\label{sec:IV}

We conclude with a discussion of some aspects of the kinetic theory and its applications as formulated in Secs.~\ref{sec:II} and
\ref{sec:III}, as well an outline of some applications and generalizations that we leave for future work.

\subsection{General aspects of the kinetic theory}
\label{subsec:IV.A}

We have considered a kinetic theory that provides an effective, or reduced, description of a fermionic many-body system
in terms of the $\mu$-space distribution function. In general, the latter depends on all of the $N$-particle distribution
functions. The Boltzmann equation closes the system by factorizing the two-particle distribution. In the
Landau-Boltzmann equation the simple product of single-particle distribution functions is generalized to account
for the Pauli principle, see Eq.~(\ref{eq:D.3}).
This procedure is controlled in the limit of low density (e.g., for a classical dilute gas), but not in general. The hydrodynamic description
in Sec.~\ref{subsec:III.B} further reduces the number of explicitly considered degrees of freedom by projecting onto the
space of the five hydrodynamic quantities. This is controlled in the limit of low frequencies and small wave numbers.
Either description involves an averaging process that ignores fluctuations of, and correlations between, the
hydrodynamic modes. The Langevin/Landau-Lifshitz concept of a fluctuating force is designed to take these into
account, and fluctuating hydrodynamics is capable of describing mode-mode coupling effects that are not included
in the averaged equations. For an example in the context of a non-equilibrium classical fluid, see the long-range correlations originally obtained by
means of kinetic theory and mode-coupling theory\cite{Kirkpatrick_Cohen_Dorfman_1982a, Kirkpatrick_Cohen_Dorfman_1982b,
Kirkpatrick_Cohen_Dorfman_1982c} that were also derived by using fluctuating hydrodynamics.\cite{Ronis_Procaccia_1982}

A remarkable aspect of LFL theory is the fact that the entropy has the same functional form as in a noninteracting Fermi gas, see Eq.~(\ref{eq:D.5a})  
and Eq.~(A.11) in Paper I. This is usually motivated by the assumption that the classification of energy levels is the same with or without
interactions.\cite{Landau_Lifshitz_IX_1991} An equivalent statement is that the equilibrium single-particle distribution in a Fermi liquid has the same
functional form as in a non-interacting Fermi gas, i.e., it is given by the Fermi-Dirac distribution. The equivalency follows from the
observation that maximizing the entropy subject to the constraints of the conservation of particle number and energy must yield
the equilibrium distribution. An argument for the single-particle distribution having the Fermi-Dirac form can be made as follows.
The usual derivation of the Fermi-Dirac distribution within Statistical Mechanics considers the grand canonical potential of the
particles in a given quantum state with energy $\epsilon_p$ that can exchange particles with all other states.\cite{Landau_Lifshitz_V_1980}
In the case of fermions the state can be occupied by at most one particle. Therefore, the energy of $N_p$ particles in that state
is $N_p \epsilon_p$ (i.e., either zero or $\epsilon_p$) whether or not the fermions in the system as a whole interact. (Note that this
is {\em not} true for bosons.) Therefore, the derivation of the average occupation number, i.e., the Fermi-Dirac distribution, is
unchanged from the noninteracting case. The energy $\epsilon_p$ depends on the interactions in a complicated way, but the
distribution function depends on them only via $\epsilon_p$. This fixes the functional form of the entropy up to a constant, and
 the latter is determined by the Third Law.

\subsection{Outlook}
\label{subsec:IV.B}
 
We finally list some problems that would be interesting to investigate using the formalism developed in this paper.      

\begin{enumerate}[leftmargin=10pt]
\item
The static $\mu$-space response function given in Eq.~(\ref{eq:2.17}), which we have used to specify the Langevin force
operator correlations, does not include long-range correlations that are present in certain static response functions in the quantum
limit. For example, it is known that the static spin susceptibility $\chi_s({\bm k})$ is a nonanalytic function of the wave number, which
reflects long-range correlations in the zero-temperature quantum system,\cite{Belitz_Kirkpatrick_Vojta_1997} and this long-range behavior
fundamentally changes the nature of some quantum phase transitions.\cite{Brando_et_al_2016a} The static number density and
momentum response functions, on the other hand, do not display this long-ranged behavior.\cite{Belitz_Kirkpatrick_Vojta_2002}       
These effects will become relevant once the spin degrees of freedom, which we have not considered, are included in the theory. 

\item As already mentioned, in classical statistical mechanics it is well known that in non-equilibrium steady states equal-time
correlation functions and static response functions are generically of extraordinarily long range; see Refs.~\onlinecite{Dorfman_Kirkpatrick_Sengers_1994,
Dorfman_vanBeijeren_Kirkpatrick_2021} and references therein. They also display a well defined kind of generalized rigidity.\cite{Kirkpatrick_Belitz_Dorfman_2021}
It would be of great interest to consider quantum systems in a corresponding non-equiblibrium state and examine the same
functions in both the collisionless regime and the hydrodynamic regime. 

\item In order to derive the fluctuating hydrodynamic equations for a disordered fermion system in Sec.~\ref{subsec:III.C} we assumed
that the disorder average has already been performed as a first step. An interesting extension would be to derive hydrodynamic
equations for a fixed impurity configuration. This could be done by working in real space and taking the fermion-impurity
collision operator, Eq.~(\ref{eq:A.2}), to be proportional to a fixed impurity density $n_i({\bm x})$. The hydrodynamic
equations would then implicitly depend on $n_i({\bm x})$ via a dependence of the transport coefficients on $n_i({\bm x})$.
In order to derive physical quantities one would then have to average over both the noise fluctuations and the static disorder
fluctuations.

\item Another aspect of the model considered in Sec.~\ref{subsec:III.C} is that both fermion number and energy are
conserved in collisions. This is the case, for instance, if fermion-fermion scattering and fermion-impurity scattering
are present. Another interesting class of problems conserves only the fermion number; this can be realized, for instance,
by adding electron-phonon scattering. At the level of the hydrodynamic description the only soft mode would then be
the diffusive number density mode. The relevant equation can be derived by using the techniques from Sec.~\ref{subsec:III.B}
and \ref{subsec:III.C} and projecting on the mode $a_1$ only. In order to describe heat transport in such a system one would
have to consider the energy transport in the phonon system as well as the energy carried by the electrons and take into
account that the energy in the combined system is still conserved.

\end{enumerate}

\appendix
\section{Some linearized collision operators}
\label{app:A}

In Appendix C of Paper I we gave expressions for collision operators describing fermion-fermion scattering, fermion-phonon
scattering, and fermion-impurity scattering respectively. They remain valid in the present context if we replace the distribution
function $\phi$ by the operator-valued function $\hat\phi$, and we list them here again for completeness. Remarks concerning
their derivations from the more general expressions given in Ref.~\onlinecite{Landau_Lifshitz_X_1981} can be found in Paper I.

The linearized fermion-fermion collision operator can be written
\begin{widetext}
\bse
\label{eqs:A.1}
\bea
\Lambda_{\text{f-f}}({\bm p})\,\hat\phi({\bm p}) &=& \frac{1}{1-f_{\text{eq}}({\bm p})}\,\frac{1}{V^3} \sum_{{\bm p}'\!,{\bm p}_1,{\bm p}_1'}
     W({\bm p},{\bm p}_1;{\bm p}',{\bm p}_1')\,\delta(\epsilon_p + \epsilon_{p_1} - \epsilon_{p'} - \epsilon_{p_1'})\, \delta({\bm p}+{\bm p}_1-{\bm p}'-{\bm p}_1') 
     \nonumber\\
     && \times f_{\text{eq}}({\bm p}_1) [1 - f_{\text{eq}}({\bm p}')] [1 - f_{\text{eq}}({\bm p}_1')]\,[\hat\phi({\bm p}') + \hat\phi({\bm p}_1') - \hat\phi({\bm p}) - \hat\phi({\bm p}_1)]\ . 
\label{eq:A.1a}
\eea
Here $W$ is the probability for two fermions in momentum states ${\bm p}$ and ${\bm p}_1$ to be scattered into momentum
states ${\bm p}'$ and ${\bm p}_1'$. Time reversal symmetry implies
\be
W({\bm p},{\bm p}_1;{\bm p}',{\bm p}_1') = W({\bm p}',{\bm p}_1';{\bm p},{\bm p}_1)\ .
\label{eq:A.1b}
\ee
\ese

The fermion-impurity collision operator is
\be
\Lambda_{\text{f-i}}({\bm p}) \,\hat\phi({\bm p}) =  \frac{1}{V} \sum_{{\bm p'}} W({\bm p}',{\bm p})\,\delta(\epsilon_{p'} - \epsilon_p) \left[\hat\phi({\bm p}') - \hat\phi({\bm p})\right]\ ,
\label{eq:A.2}
\ee
with $W$ another scattering probability. Here the fermion particle number and energy are conserved, but the momentum is not.

The fermion-boson collision operator can be written
\bea
\Lambda_{\text{f-b}}({\bm p})\,\hat\phi({\bm p}) &=&  \frac{1}{f_{\text{eq}}({\bm p})(1-f_{\text{eq}}({\bm p}))}\,\frac{1}{V^2} \sum_{{\bm p}'\!,{\bm k}}
     \delta({\bm p}'-{\bm p}-{\bm k})\,n_{\text{eq}}({\bm k}) \Bigl[W({\bm p}';{\bm p},{\bm k})\,f_{\text{eq}}({\bm p})(1-f_{\text{eq}}({\bm p}'))\,
          \delta(\epsilon_{p'}-\epsilon_p-\omega_k)
\nonumber\\
          && \hskip 100pt + W({\bm p}',-{\bm k};{\bm p})\,f_{\text{eq}}({\bm p}')(1-f_{\text{eq}}({\bm p}))\,\delta(\epsilon_{p'}-\epsilon_p+\omega_k) \Bigr]
               \left[\hat\phi({\bm p}') - \hat\phi({\bm p})\right]\ .
\label{eq:A.3}
\eea
\end{widetext}
Here $n_{\text{eq}}({\bm k}) = 1/(\exp(\omega_k/T)-1)$ is the equilibrium Bose-Einstein distribution function, $\omega_k$ is the energy
of a boson with wave number $k$, and $W$ is again a transition probability. The fermion particle number is still conserved, but the fermion 
momentum and energy are not. Umklapp processes can be taken into account if desirable. 

Finally, a quantum version of the Bhatnagar-Gross-Krook collision operator in classical kinetic 
theory\cite{Bhatnagar_Gross_Krook_1954, Dorfman_vanBeijeren_Kirkpatrick_2021} is useful for explicit calculations it 
uses a single relaxation time $\tau(T)$ and builds in the five conservation laws for the particle number, momentum, and energy.
It is given by
\bse
\label{eqs:A.4}
\be
\Lambda_{\text{f-f}}^{\text{BGK}}({\bm p}) = \frac{-1}{\tau}\,{\cal P}_{\perp}\ ,
\label{eq:D.5a}
\ee
where
\be
{\cal P}_{\perp} = \openone -  \sum_{\alpha=1}^5 \frac{\vert a_{\alpha}({\bm p}\rangle\langle a_{\alpha}({\bm p})\vert}{\langle a_{\alpha}({\bm p})\vert a_{\alpha}({\bm p})\rangle}
\label{eq:A.4b}
\ee
\ese
is the projection operator defined in Sec.~\ref{subsec:III.B} that enforces the five conservation laws.

In general, the linearized collision operator $\Lambda({\bm p})$ is a sum of several of these individual collision operators.

\section{The $\mu$-space static response function, and the Langevin force correlation}
\label{app:B}

In this appendix we show how to derive Eqs.~(\ref{eqs:2.15}). 

We can obtain a kinetic equation for the dynamic response function by adding a term $\delta\mu({\bm p},{\bm x},t)$
to the chemical potential that depends on momentum, space, and time, and acts as an external field conjugate to the
operator $\delta{\hat f}$. The single-particle energy from Eq.~(\ref{eq:2.5}) then becomes
\be
{\hat\epsilon}({\bm p},{\bm x}) = \epsilon_p + \delta{\hat\epsilon}({\bm p},{\bm x}) - \delta\mu({\bm p},{\bm x},t)\ .
\label{eq:B.1}
\ee
In the kinetic equation this produces a new term, and instead of Eq.~(\ref{eq:2.6b}) we obtain
\begin{widetext}
\be
\left[-i\omega + L_{\bm k}({\bm p})\right] w({\bm p}) {\hat\phi}({\bm p},{\bm k},\omega) - i{\bm k}\cdot{\bm v}_p\, \delta{\tilde\mu}({\bm p},{\bm k},\omega) 
     = {\hat F}_{\text{L}}({\bm p},{\bm k},\omega)\ ,
\label{eq:B.2}
\ee
where $\delta{\tilde\mu}({\bm p},{\bm k},\omega) = w({\bm p}) \delta\mu({\bm p},{\bm k},\omega)$. The dynamic 
response function infinitesimally above the real frequency axis is given by the functional derivative
\be
\chi({\bm p}_1,{\bm p}_2;{\bm k},\omega + i\epsilon) = \delta\langle{\hat\phi}({\bm p}_1,{\bm k},\omega)\rangle/\delta{\tilde\mu}({\bm p}_2,{\bm k},\omega)\ .
\label{eq:B.3}
\ee
Equation (\ref{eq:B.2}) yields
\bse
\label{eqs:B.4}
\be
\left[-i\omega + L_{\bm k}({\bm p}_1)\right] \chi({\bm p}_1,{\bm p}_2;{\bm k},\omega + i\epsilon) = V \delta_{{\bm p}_1,{\bm p}_2}\,\frac{i{\bm k}\cdot{\bm v}_{p_1}}{w({\bm p}_1)}\ ,
\label{eq:B.4a}
\ee
which has the formal solution
\be
\chi({\bm p}_1,{\bm p}_2;{\bm k},\omega + i\epsilon) = - \left[ \omega + iL_{\bm k}^{(1)}({\bm p}_1) - i\Lambda({\bm p}_1) \right]^{-1}\,V \delta_{{\bm p}_1,{\bm p}_2}\,
     \frac{{\bm k}\cdot{\bm v}_{p_1}}{w({\bm p}_1)}\ ,
\label{eq:B.4b}
\ee
\ese
\end{widetext}
with $L_{\bm k}^{(1)}$ from Eq.~(\ref{eq:2.6d}). Using Eq.~(\ref{eq:2.9b}) in (\ref{eq:2.10}), and performing the frequency integration before taking the imaginary part
(note that $iL_{\bm k}^{(1)}$ is real, whereas $i\Lambda$ is imaginary), yields 
\bse
\label{eqs:B.5}
\be
   %
\chi({\bm p}_1,{\bm p}_2;{\bm k}) = - \Im \left(L_{\bm k}({\bm p}_1)\right)^{-1} V  \delta_{{\bm p}_1,{\bm p}_2}\, \frac{{\bm k}\cdot{\bm v}_{p_1}}{w({\bm p}_1)}\ .
\label{eq:B.5a}
\ee
Alternatively we can obtain the static response by putting $\omega=0$ in Eq.~(\ref{eq:B.4b}),
\be
\chi({\bm p}_1,{\bm p}_2;{\bm k}) =  \left(L_{\bm k}({\bm p}_1)\right)^{-1} V  \delta_{{\bm p}_1,{\bm p}_2}\, \frac{i {\bm k}\cdot{\bm v}_{p_1}}{w({\bm p}_1)}\ .
\label{eq:B.5b}
\ee
\ese
Equations~(\ref{eq:B.5a}) and (\ref{eq:B.5b}) are equivalent due to the spectral representation of the response function,
see Eqs.~(\ref{eqs:2.9}) and (\ref{eq:2.10}). 
Since the static response cannot depend on transport coefficients,\cite{static_response_footnote} the collision operator implicit in this expression cannot contribute
and we can evaluate Eqs.~(\ref{eqs:B.5}) for $\Lambda({\bm p}_1) = -i\epsilon$. The static response function thus obeys an integral equation
\be
L_{\bm k}^{(1)}({\bm p}_1)\, \chi({\bm p}_1,{\bm p}_2;{\bm k}) = V \delta_{{\bm p}_1,{\bm p}_2}\, \frac{i {\bm k}\cdot{\bm v}_{p_1}}{w({\bm p}_1)}\ .
\label{eq:B.6}
\ee
With $L_{\bm k}^{(1)}$ from Eqs.~(\ref{eqs:2.6}) this is Eq.~(\ref{eq:2.15b}). 

We still need to show how to obtain Eq.~(\ref{eq:2.15a}) from (\ref{eq:2.14}). To this end we write
\be
\chi({\bm p}_1,{\bm p}_2;{\bm k}) = V \delta_{{\bm p}_1,{\bm p}_2}\,\frac{1}{w({\bm p}_1)} + \delta\chi({\bm p}_1,{\bm p}_2;{\bm k})
\label{eq:B.7}
\ee
From Eq.~(\ref{eq:2.15b}) we see that $\delta\chi$ obeys the integral equation
\bse
\label{eqs:B.8}
\be
\left[1 + I({\bm p}_1)\right] \delta\chi({\bm p}_1,{\bm p}_2;{\bm k}) = \frac{-1}{N_0}\,F({\bm p}_1,{\bm p}_2)\ ,
\label{eq:B.8a}
\ee
with $I({\bm p})$ from Eq.~(\ref{eq:2.6e}). 
Since $\delta\chi$ and $F$ are both symmetric under interchange of the two ${\bm p}$-variables, Eq.~(\ref{eq:B.8a}) can also be written
\be
\left[1 + I({\bm p}_2)\right] \delta\chi({\bm p}_1,{\bm p}_2;{\bm k}) = \frac{-1}{N_0}\,F({\bm p}_1,{\bm p}_2)\ .
\label{eq:B.8b}
\ee
\ese
Simple algebra that makes use of Eqs.~(\ref{eqs:B.8}) then yields
\be
\left[L_{\bm k}^{(1)}({\bm p}_1) + L_{-{\bm k}}^{(1)}({\bm p}_2)\right] \chi({\bm p}_1,{\bm p}_2;{\bm k}) = 0.
\label{eq:B.9}
\ee
Equation~(\ref{eq:2.14}) is therefore equivalent to Eq.~(\ref{eq:2.15a}).

\section{The $\mu$-space distribution, and the hydrodynamic modes}
\label{app:C}

In order to illustrate the connection between the approaches in Paper I and the current paper, we consider the kinetic equation (\ref{eq:2.6c})
and its formal solution
\be
\vert{\hat\phi}({\bm p},{\bm k},\omega)\rangle = \left[-i\omega + L_{\bm k}({\bm p})\right]^{-1} \vert{\hat F}_{\text{L}}({\bm p},{\bm k},\omega)\rangle\ . 
\label{eq:C.1}
\ee
Now we insert a complete set of modes between the propagator and the fluctuating force on the right-hand side. 
As part of that complete set we choose the five hydrodynamic modes $\psi_{\alpha}({\bm k},{\bm p})$ ($\alpha = 1,\ldots, 5$) 
that are defined as the eigenfunctions of the kinetic operator $L_{\bm k}({\bm p})$ with eigenvalues $\omega_{\alpha}({\bm k})$,
see Eq.~(3.1) in Paper I. For the part of ${\hat\phi}$ in the space spanned by the hydrodynamic modes we then have (note that
the left and right eigenfunctions are not identical, see Paper I)
\begin{widetext}
\be
\vert{\hat\phi}_{\text{hyd}}({\bm p},{\bm k},\omega)\rangle = \sum_{\alpha} \,\vert\psi_{\alpha}^{(\text{R})}({\bm k},{\bm p})\rangle\, \left[-i\omega + \omega_{\alpha}({\bm k})\right]^{-1} 
     \langle\psi_{\alpha}^{(\text{L})}({\bm k},{\bm p})\vert {\hat F}_{\text{L}}({\bm p},{\bm k},\omega)\rangle
     \ .
\label{eq:C.2}
\ee
\end{widetext}
Here we have assumed that the hydrodynamic modes are normalized to unity. Equation~(\ref{eq:C.2}) is equivalent
to an explicit solution of the hydrodynamic equations in Sec.~\ref{subsec:III.B}. The eigenfrequencies $\omega_{\alpha}({\bm k})$
and the eigenfunctions $\psi_{\alpha}$ have been determined explicitly in Paper I. To zeroth order in an expansion of the hydrodynamic modes in 
powers of the wave number $k = \vert{\bm k}\vert$, the $\psi_{\alpha}$ are
linear combinations of the zero eigenfunctions $a_{\alpha}({\bm p})$ of the collision operator, see Eq.~(3.6) in Paper I, and therefore have no overlap with the
fluctuating force. However, their contributions to first order in the $k$-expansion are elements of the complementary space ${\cal L}_{\perp}$, see
Eqs.~(3.28) in Paper I. The leading contribution to the right-hand side of Eq.~(\ref{eq:C.2}) is thus of $O(k)$. This is consistent with the fact
that the Langevin forces in fluctuating hydrodynamics are currents, see Sec.~\ref{subsubsec:III.B.2} and Ref.~\onlinecite{Landau_Lifshitz_VI_1959}.

\section{$H$-theorem for the nonlinear Boltzmann-Landau kinetic equation}
\label{app:D}

In this appendix  we derive an $H$-theorem for the fermionic quantum kinetic equation; see ch. 10.3.3 in Ref.~\onlinecite{Dorfman_vanBeijeren_Kirkpatrick_2021} for
a closely related discussion. Let $f_{\bm p}({\bm x},t) = \langle{\hat f}({\bm p},{\bm x},t)\rangle$
be the average non-equilibrium distribution function, $\delta f_{\bm p} = f_{\bm p} - f_{\text{eq}}({\bm p})$ its deviation from the equilibrium distribution, 
and 
\be
\epsilon_{\bm p}({\bm x},t) = \langle{\hat\epsilon}({\bm p},{\bm x},t)\rangle = \epsilon_p + \frac{1}{N_0V}\sum_{{\bm p}'} F({\bm p},{\bm p}')\,\delta f_{{\bm p}'}({\bm x},t)
\label{eq:D.1}
\ee
the average single-particle energy. The nonlinear Boltzmann-Landau kinetic equation for $f_{\bm p}$ is\cite{Landau_Lifshitz_X_1981}
\be
\partial_t f_{\bm p} + ({\bm\nabla_{\bm x}} f_{\bm p})\cdot{\bm\nabla}_{\bm p}\epsilon_{\bm p} - ({\bm\nabla}_{\bm p} f_{\bm p})\cdot{\bm\nabla}_{\bm x}\epsilon_{\bm p} = C(f)
\label{eq:D.2}
\ee
For the collision integral we assume the one appropriate for fermion-fermion collisions (Ref.~\onlinecite{Landau_Lifshitz_X_1981}, see Eq.~(\ref{eq:A.1a}) for the
linearized version)
\begin{widetext}
\bea
C(f) &=& \frac{1}{V^3} \sum_{{\bm p}'\!,{\bm p}_1,{\bm p}_1'}  
     W({\bm p},{\bm p}_1;{\bm p}',{\bm p}_1')\,\delta(\epsilon_p + \epsilon_{p_1} - \epsilon_{p'} - \epsilon_{p_1'})\, \delta({\bm p}+{\bm p}_1-{\bm p}'-{\bm p}_1') 
\nonumber\\
&& \times\left[ f_{{\bm p}'} f_{{\bm p}_1'} (1 - f_{\bm p})(1 - f_{{\bm p}_1}) - f_{\bm p} f_{{\bm p}_1} (1 - f_{{\bm p}'})(1 - f_{{\bm p}_1'})\right]\ .
\label{eq:D.3}
\eea
\end{widetext}
The transition rate $W$ is positive and has the symmetry properties
\bse
\label{eqs:D.4}
\bea
W({\bm p},{\bm p}_1;{\bm p}',{\bm p}_1') &=& W(-{\bm p},-{\bm p}_1;-{\bm p}',-{\bm p}_1')
\label{eq:D.4a}\\
&=& W({\bm p}',{\bm p}_1';{\bm p},{\bm p}_1)
\label{eq:D.4b}\\
&=& W({\bm p}_1,{\bm p};{\bm p}_1',{\bm p}')\ ,
\label{eq:D.4c}
\eea
\ese
which express invariance under spatial inversions, time reversal, and interchange of particles.
The local $h$-function is usually defined as minus the entropy density
\bse
\label{eqs:D.5}
\be
h({\bm x},t) = \frac{1}{V}\sum_{\bm p} \left[ f_{\bm p}\ln f_{\bm p}  + (1 - f_{\bm p}) \ln (1 - f_{\bm p})  \right]\ .
\label{eq:D.5a}
\ee
The global $H$-function is
\be
H(t) = \int d{\bm x}\,h({\bm x},t)\ .
\label{eq:D.5b}
\ee
\ese
The $H$-theorem consists of the statement
\be
dH/dt \leq 0\ ,
\label{eq:D.6}
\ee
with the equal sign holding if and only if $f_{\bm p}$ is the equilibrium distribution $f_{\text{eq}}({\bm p})$, i.e., the Fermi-Dirac distribution.

In order to prove the $H$-theorem we start by using Eq.~(\ref{eq:D.2}) to show that the local $h$-function obeys a continuity equation
\bse
\label{eqs:D.7}
\be
\partial_t h + {\bm\nabla}_{\bm x}\cdot{\bm j}_h = I(f)\ ,
\label{eq:D.7a}
\ee
with
\be
{\bm j}_h({\bm x},t) =  \frac{1}{V}\sum_{\bm p} ({\bm\nabla}_{\bm p}\epsilon_{\bm p})\left[ f_{\bm p}\ln f_{\bm p}   + (1 - f_{\bm p})\ln (1 - f_{\bm p})  \right]
\label{eq:D.7b}
\ee
the $h$-current density, and
\be
I(f) = \frac{1}{V}\sum_{\bm p} C(f) \ln\left(\frac{f_{\bm p}}{1 - f_{\bm p}}\right)
\label{eq:D.7c}
\ee
\ese
a source term. To arrive at this result we have integrated by parts in momentum space and assumed that
any surface terms vanish. One now can show that $I(f)\leq 0$ by proceeding as in the case of the classical
Boltzmann equation:\cite{Cercignani_1988} Substitute Eq.~(\ref{eq:D.3}) into Eq.~(\ref{eq:D.7c}), interchange
${\bm p}$ and ${\bm p}_1$ as well as ${\bm p}'$ and ${\bm p_1}'$, add the two expressions and divide by $2$,
using Eq.~(\ref{eq:D.4c}). Next interchange ${\bm p}$ and ${\bm p}'$ as well as ${\bm p}_1$ and ${\bm p}_1'$,
again add the two expressions and divide by $2$, using Eq.~(\ref{eq:D.4b}). The result is
\begin{widetext}
\bea
I(f) &=& \frac{-1}{4 V^4} \sum_{{\bm p},{\bm p}_1,{\bm p}',{\bm p}_1'}  W({\bm p},{\bm p}_1;{\bm p}',{\bm p}_1')\,\delta(\epsilon_p + \epsilon_{p_1} - \epsilon_{p'} - \epsilon_{p_1'})\, \delta({\bm p}+{\bm p}_1-{\bm p}'-{\bm p}_1') 
\nonumber\\
&&\times f_{\bm p} f_{{\bm p}_1} (1 - f_{{\bm p}'})(1 - f_{{\bm p}_1'}) \left[\frac{f_{{\bm p}'} f_{{\bm p}_1'} (1 - f_{\bm p})(1 - f_{{\bm p}_1})}{ f_{\bm p} f_{{\bm p}_1} (1 - f_{{\bm p}'})(1 - f_{{\bm p}_1'})} - 1\right] \ln\left(\frac{f_{{\bm p}'} f_{{\bm p}_1'} (1 - f_{\bm p})(1 - f_{{\bm p}_1})}{ f_{\bm p} f_{{\bm p}_1} (1 - f_{{\bm p}'})(1 - f_{{\bm p}_1'})}\right)
\label{eq:D.8}
\eea
\end{widetext}
But
\be
(x-1)\ln x \geq 0 \qquad \forall x>0
\label{eq:D.9}
\ee
with the equal sign holding if and only if $x=1$ (this is a special case of Jensen's inequality). Accordingly, $I(f) \leq 0$. Integrating
Eq.~(\ref{eq:D.7a}) over all of space, and again ignoring surface terms, yields Eq.~(\ref{eq:D.6}).

It remains to be shown that $dH/dt = 0$ if and only if $f_{\bm p} = f_{\text{eq}}({\bm p})$. This follows in exact analogy to the classical
case:\cite{Cercignani_1988} Suppose $f_{\bm p}$ is the Fermi-Dirac
distribution, Eq.~(\ref{eq:2.1b}). Then $\ln[f_{\bm p}/(1-f_{\bm p})] = -\xi_{\bm p}/T$ is a collision invariant, and hence $I(f_{\text{eq}}) = 0$.
Now suppose $I(f) = 0$. Then by Eq.~(\ref{eq:D.9}) $\ln[f_{\bm p}/(1-f_{\bm p})]$ must be a collision invariant, and thus a linear
combination of the five conserved quantities $\epsilon_p$ (energy), $1$ (particle number), and ${\bm p}$ (momentum). The latter
has a zero coefficient by Galilean invariance, and therefore $\ln[f_{\bm p}/(1-f_{\bm p})] =-\epsilon_p/T + \mu$, with minus the
inverse temperature and the chemical potential serving as the coefficients in the linear combination. $I(f)=0$ thus implies that
$f$ is given by Eq.~(\ref{eq:2.1b}).

\section{Some useful identities}
\label{app:E}

In this appendix we list, for completeness, some thermodynamic identities that were derived in Paper I and that we have used throughout 
the main text.

Two common averages are, see Eqs.~(A19a) and (A23a) in Paper I,
\bse
\label{eqs:E.1}
\bea
\langle v_p^i\vert p^j\rangle &=& \delta^{ij} n\ ,
\label{eq:E.1a}\\
\langle v_p^i\vert p^j a_5({\bm p})\rangle &=& \delta^{ij} T(\partial p/\partial T)_{N,V}\ .
\label{eq:E.1b}
\eea
\ese

The average energy density $e$, the average pressure $p$, the average density $n$, and the temperature derivative of the
pressure are related by
\be
n\langle\epsilon_p\rangle_w + T(\partial p/\partial T)_{N,V} = e + p\ .
\label{eq:E.2}
\ee
For a derivation, see Eq.~(3.16) in Paper I. Another relation between $\langle\epsilon_p\rangle_w$ and a thermodynamic
derivative is
\be
\langle\epsilon_p\rangle_w = \mu - T(\partial\mu/\partial T)_{N,V}\ ,
\label{eq:E.3}
\ee
see Eq.~(A7) in Paper I.

\section{Fluctuating Navier-Stokes equations for a clean metal}
\label{app:F}

In this appendix we briefly discuss the changes that occur in Sec.~\ref{subsec:III.B} if the kinetic theory is applied to
the electron fluid in a metal, i.e., a fermion system with a long-ranged Coulomb interaction. We will assume that the
metal is clean in the sense that there is no electron-impurity scattering.

A long-ranged Coulomb interaction results in an additional contribution $4\pi N_0 e^2/k^2$ to the Landau interaction 
parameter $F_0$. In the fluid velocity equation (\ref{eq:3.20a}) this leads to an additional term
\begin{equation*}
i k\,\frac{4\pi n e^2}{k^2}\,\delta{\hat n}({\bm k},\omega)\ .
\end{equation*}
Combined with the density equation (\ref{eq:3.19}) this singular term leads to nonzero eigenfrequencies 
$\omega = \pm\omega_p + O(k^2)$ in the longitudinal (parallel to ${\bm k}$) channel, with $\omega_p = (4\pi ne^2/m)^{1/2}$
the plasmon frequency. This in turn implies that the frequency in Eq.~(\ref{eq:3.16}) can no longer be neglected
compared to the collision operator, as both are of $O(1)$ with respect to $k$. 

In order to capture these effects it is advantageous to introduce, instead of Eq.~(\ref{eq:3.11b}), longitudinal and
transverse momentum modes as
\bse
\label{eqs:F.1}
\bea
a_2({\bm p}) &=& {\hat{\bm k}}\cdot{\bm p}\ ,
\label{eq:F.1a}\\
a_3({\bm p}) &=& {\hat{\bm k}_{\perp}^{(1)}}\cdot{\bm p}\ ,
\label{eq:F.1b}\\
a_4({\bm p}) &=& {\hat{\bm k}_{\perp}^{(2)}}\cdot{\bm p}\ ,
\label{eq:F.1b]c}\
\eea
\ese
where $\hat{\bm k}$ and $ {\hat{\bm k}}_{\perp}^{(1,2)}$ are three pairwise orthogonal unit vectors in the direction
of ${\bm k}$ and the two directions perpendicular to ${\bm k}$, respectively. Repeating the procedure from 
Sec.~\ref{subsubsec:III.B.2} we find that the transverse fluid velocity equation is unchanged, and the transverse
stress tensor is still given in terms of a transverse shear viscosity that is unchanged from the short-range case, Eq.~(\ref{eq:3.20c}),
\begin{widetext}
\bse
\label{eqs:F.2}
\be
\eta_{\perp} = -\left\langle(\hat{\bm k}_{\perp}\cdot{\bm p})(\hat{\bm k}\cdot{\bm v}_p) \big\vert \Lambda^{-1}({\bm p}) \big\vert 
      (\hat{\bm k}\cdot{\bm v}_p)(\hat{\bm k}_{\perp}\cdot{\bm p})\right\rangle\ .
\label{eq:F.2a}
\ee
The dissipative term in the equation for the longitudinal fluid velocity, on the other hand, is given in terms of
\be
\eta_{\parallel}^{\pm} = -\left\langle(\hat{\bm k}_{\perp}\cdot{\bm p})(\hat{\bm k}\cdot{\bm v}_p) \big\vert (\pm i\omega_p + \Lambda({\bm p}))^{-1} \big\vert 
      (\hat{\bm k}\cdot{\bm v}_p)(\hat{\bm k}_{\perp}\cdot{\bm p})\right\rangle\ .
\label{eq:F.2b}
\ee
\ese
\end{widetext}
This transport coefficient, which can be interpreted as a high-frequency shear viscosity, has both a real part
and an imaginary part, and thus contributes both to the ${\bm k}$-dependence of the plasmon frequency
and to the plasmon damping, see Paper I. 

Similarly, the fluctuating stress tensor, Eq.~(\ref{eq:3.20h}), has transverse parts 
\bse
\label{eqs:F.3}
\be
\left({\hat\tau}_{\text{L}}\right)_{\perp}^{(1,2)}({\bm x},\omega) = \left\langle ({\hat{\bm k}}_{\perp}^{(1,2)}\cdot{\bm p})({\hat{\bm k}}\cdot{\bm p})\vert\Lambda^{-1}({\bm p})\vert {\hat F}_{\text{L}}({\bm p},{\bm x},\omega)\right\rangle
\label{eq:F.3a}
\ee
and a longitudinal part
\be
\left({\hat\tau}_{\text{L}}\right)_{\parallel}^{\pm}({\bm x},\omega) = \left\langle ({\hat{\bm k}}\cdot{\bm p})^2\vert(\pm i \omega_p +\Lambda^{-1}({\bm p}))^{-1}\vert {\hat F}_{\text{L}}({\bm p},{\bm x},\omega)\right\rangle
\ee
\ese
These different contributions to the fluctuating force are not correlated with each other. The transverse noise correlation is given
in terms of the ordinary zero-frequency shear viscosity, as in Eq.~(\ref{eq:3.24a}), while the longitudinal correlation depends on the
dissipative (i.e., real) part of the high-frequency shear viscosity, Eq.~(\ref{eq:F.2b}).

We finally mention that in a two-dimensional metal the Coulomb contribution to the interaction parameter $F_0$ is proportional to $1/k$,
and consequently the plasmons are soft with a frequency $\omega_p \propto k^{1/2}$. As a result, the longitudinal and transverse
shear viscosities in the long-wavelength limit are identical. 



\begin{thebibliography}{44}
\bibitem[{the()}]{thermodynamic_limit_footnote}
\bibinfo{note}{Throughout this paper we assume the thermodynamic limit,
  $V\to\infty$, so $(1/V)\sum_{\bm p}$ should be considered as synonymous with
  $\int d{\bm p}/(2\pi)^d$ in $d$ spatial dimensions.}
\bibitem[{ave()}]{average_footnote}
\bibinfo{note}{The subscript $w$ distinguishes this average from the one
  defined after Eq.~(\ref{eq:2.1a}). The scalar product, Eq.~(\ref{eq:2.3a}),
  also uses an angular bracket notation, but there can be no confusion between
  it and either average.}
\bibitem[{dyn()}]{dynamics_footnote}
\bibinfo{note}{$\hat\phi$ is defined in terms of the Heisenberg operators
  ${\hat a}^{\dagger}$ and ${\hat a}$, and hence its time evolution is given by
  the Heisenberg equations with the full Hamiltonian of the many-body system.
  The idea behind the kinetic theory is to replace this exact, but entirely
  formal, time evolution by a kinetic equation with a fluctuating force. See
  Sec.~\ref{sec:IV} for further discussion of this point.}
\bibitem[{Lan()}]{Landau_parameters_footnote}
\bibinfo{note}{The integral equation (\ref{eq:2.15b}) remains separable if all
  terms in the expansion in Legendre polynomials are included. If desirable,
  $\chi({\bm p}_1,{\bm p}_2;{\bm k})$ can thus be given in closed form with all
  Landau parameters $F_{\ell}$ included.}
\bibitem[{col()}]{collision_operator_footnote}
\bibinfo{note}{In general the collision operator also has a wave number and
  frequency dependence that we neglect.}
\end{thebibliography}

\end{document}